\documentclass[floatfix,%
 aip, jcp,
 amsmath,amssymb,
 reprint,%‪./ActaMaterialia.tex, 917‬
]{revtex4-1}

\usepackage{graphicx}% Include figure files
\usepackage{dcolumn}% Align table columns on decimal point
\usepackage{bm}% bold math
\usepackage{longtable}
\usepackage[utf8]{inputenc}
\DeclareUnicodeCharacter{200A}{!!!FIX ME!!!} 
\usepackage[T1]{fontenc}
\usepackage{mathptmx}
\usepackage{etoolbox}

\makeatletter
\def\@email#1#2{%
 \endgroup
 \patchcmd{\titleblock@produce}
  {\frontmatter@RRAPformat}
  {\frontmatter@RRAPformat{\produce@RRAP{*#1\href{mailto:#2}{#2}}}\frontmatter@RRAPformat}
  {}{}
}%
\makeatother
\begin{document}

\preprint{AIP/123-QED}

\title{On the  origin of in-gap states in amorphous Ge$_2$Sb$_2$Te$_5$}

\author{Omar Abou El Kheir}
\author{Marco Bernasconi}%
\email{marco-bernasconi@unimib.it}
\affiliation{ 
Department of Materials Science, University of Milano-Bicocca, Via R. Cozzi 55, I-20125 Milano, Italy
}

\date{\today}
\begin{abstract}
The localized states in the band gap of amorphous phase change alloys like Ge$_2$Sb$_2$Te$_5$ control the electrical conduction via the Poole-Frenkel mechanism. Understanding the origin of in-gap states and  their evolution in time during aging of the glass is therefore important for the control of the resistance drift in phase change memory devices. Here, we use a machine learning interatomic potential to generate several models of Ge$_2$Sb$_2$Te$_5$ whose electronic structure is then analyzed within density functional theory with a hybrid functional. A detailed statistical analysis of the structural motifs on which the in-gap states are localized, reveals that the vast majority of in-gap states involve wrong bonds (homopolar or Ge-Sb bonds) often accompanied by Ge in tetrahedral configurations or overcoordinated Ge and Sb atoms. Metadynamics simulations mimicking glass aging support the picture that structural relaxations  lead to the depletion of in-gap states and then to an increase of resistance. The simulations thus provide important insights for the mitigation of the resistance drift in phase change memory devices.
\end{abstract}
\maketitle
\onecolumngrid

\section{Introduction}

Chalcogenide alloys are exploited in electronic non-volatile memories named phase change memories (PCM) which rely on the fast and reversible transformation between the crystalline and amorphous phases induced by heating \cite{wuttig2007phase,noe2017phase,fantini,Cappelletti_2020,Redaelli2022}. The two phases feature a large difference in electrical resistivity that allows encoding a binary information which is read by a measurement of resistance at low bias. The writing/erasing operations require higher voltage pulses to induce either the melting of the crystal and subsequent amorphization (reset) or the recrystallization of the amorphous phase (set). The modulation of the size of the amorphous region during reset or the partial recrystallization during set also allows encoding several discrete or analog resistance levels for in-memory and neuromorphic computing \cite{kuzum2012nanoelectronic,tuma2016stochastic,sebastianNanotech}.
As the amorphous phase is metastable, it is subject to aging with time due to structural relaxations that drive the system toward a more ideal, i.e. stress and defect-free, glass. These structural changes lead to an increase in  electrical resistivity with time \cite{Boniardi2009}. This phenomenon called drift is described by a power law function $R=R_o ({t}/{t_o})^{\nu}$ where $R$ and $R_o$ are the resistance at time $t$ and $t_o$ and $\nu$ is the drift exponent \cite{IelminiDrift2007}.
We note, however, that deviations from this law  have been observed at short and long times in the binary phase change compound GeTe \cite{salinga2025}
which is the parent compound of the prototypical phase change material  Ge$_2$Sb$_2$Te$_5$ (GST) lying on the GeTe-Sb$_2$Te$_3$ tie-line.
The drift should be kept as low as possible as it affects the reliability of PCM and hinders the realization of multilevel cells and the operation of neuromorphic devices \cite{multilevel}. Although some options have been proposed to mitigate this problem, either by changing the programming protocol \cite{Papandreou} or the device architecture in the so-called projected PCMs \cite{Koelmans}, it would be desirable to control and reduce drift at the material level. To this aim a better understanding of this process at the atomistic level seems mandatory.
Several explanations have been proposed for the resistance drift over the years (see Refs. \cite{LeGalloReview,RatyReview,MaReview} for a review).

It was initially assumed that the drift arises from the relaxation of the stress upon time leading to an increase in the band gap \cite{karpov}. In fact, a lower drift is measured in eventually stress-free GST nanowires \cite{agarwal}, while  in PCMs the amorphous phase is under compressive stress due to the embedding in the denser crystalline matrix. However, it was later shown that both as-deposited amorphous GST and stress-free films gave the same drift exponent \cite{fantini_film}.  A more viable and alternative scenario was then proposed in which the drift results from relaxations of local defective structures  of the amorphous phase towards a more stable and ideal glass \cite{Ielmini2009}.  These structures give rise to Urbach tails and localized shallow or deep electronic states in the mobility gap. These states control the thermally activated sub-threshold conductivity which is typically described by the Poole-Frenkel model \cite{Ielmini2009,Ielmini_Poole_Frenkel}. This model is based on the assumption that most thermally generated holes (electrons) are trapped in localized states, i.e. in hole (electron) traps arising from in-gap states within the mobility gap. The applied electric field promotes the transfer of holes (electrons) from the traps to the valence (conduction) mobile states for low trap densities (Poole-Frenkel) or a trap-to-trap thermally assisted hopping for high trap densities (Poole).  As a result, the resistance shows an Arrhenius  behavior $R=R^*e^{-E_A/k_B T}$ with a zero field activation energy given by $E_A= E_F-E_{VB}$, where $E_F$ is the Fermi level, and $E_{VB}$ is the valence band mobility edge for a p-type semiconductor like GST \cite{Ielmini2009,Ielmini_Poole_Frenkel}. Extended Urbach tails and a large density of empty and filled states in the gap (trap states for electrons or holes),  have indeed been invoked to explain modulated photocurrent experiments in amorphous GeTe and GST \cite{Longeaud,Luckas2013}. In GST the resistance drift is due to a logarithmic increase with time of the activation energy for conduction $E_A$ \cite{Boniardi2009}, but in other phase change alloys, for example GeTe, the prefactor $R^*$ (see above) equally contribute to the drift with a power law dependence on time \cite{SalingaDrift2014}.
The widening of the band gap and the reduction of valence Urbach tail would shift the valence band  (VB) mobility edge to lower energy \cite{Lavizzari2009}, while the reduction of deep states and of the conduction band (CB) tail will shift upward the Fermi level. These effects would all concur to an increase of the activation energy   $E_A$ for a p-type disordered semiconductor \cite{Ielmini2009}. 
In the scenario in which the drift is due to structural relaxations, defects with the lowest activation barriers relax first. Numerical simulations based on a phenomenological Poole-Frenkel model reproduced the logarithmic increase with time of $E_A$ ($\Delta E_A \propto log \, t$) by assuming a suitable distribution  of energy barriers for structural relaxations and a suitable dependence of the activation energy for conduction on the density of in-gap states \cite{Lavizzari2009}. This picture can also explain the observed acceleration of the drift by an applied electric field by assuming that the field induces a reduction of the activation barrier for structural relaxations \cite{FantiniCampoE}.  
Instead of a distribution of activation energies that get eroded with time, the kinetics of structural relaxations was described in Ref. \cite{LeGallo2018} by a single activation energy for collective transformations that change with time due to aging.

 A correlation between drift, the widening of the band gap and the reduction of  Urbach tails was indeed provided  by optical ellipsometry measurements on GST \cite{fantini_optical}.  The measurement of $\varepsilon_{\infty}$ by Fourier transform infrared spectroscopy (FTIR) also confirmed the widening of the band gap with time \cite{Rutten2015}. In another study on a-GeTe, a widening of the optical gap was observed upon ageing, but not a change in the Urbach tail \cite{Luckas2013}. On the other hand, modulated photocurrent experiments showed a decrease in the density of deep in-gap states, but an increase in the density of shallow states close to the VB \cite{Luckas2013}. It was also shown by differential scanning calorimetry that the resistance drifts simultaneously with a release of enthalpy due to structural relaxations of the glass \cite{Wuttig2024}.
 
More recently, it was proposed that drift can arise from thermal detrapping of carriers from localized states in the gap and not from the removal of these states by structural relaxations \cite{Elliott2020}. This scenario is based on the assumption that carriers follow a non-equilibrium distribution after the reset process due to a very fast cooling. This scenario is supported by experimental evidence of a reduction of the drift by photoexcitation, even at very low temperatures, which seems more consistent with a change in the population of trap states than to an enhancement of structural relaxations \cite{Khan2020}.
Moreover, it has also been recently proposed that resistance drift originates from a change with time of the Schottky barrier height for hole injection at the interface \cite{Yalon}. Anyway,  the nature and abundance of the localized states in the mobility gap play a crucial role in these two latter scenarios as well. 
 
Experimental information on the nature of in-gap states in GST came 
from Ge K-edge x-ray absorption near-edge structure (XANES) spectrum
\cite{kolobovdrift}.  It was shown that the drift was correlated with the reduction of a step-like feature in the pre-edge XANES spectra  that was assigned to tetrahedrally coordinated Ge with the aid of electronic structure calculations \cite{kolobovdrift}.  Only a  minority fraction of Ge atoms displays a tetrahedral bonding geometry in GST or GeTe, as most atoms are in a pyramidal or defective octahedral configurations \cite{caravati2007coexistence,akola2007structural,elliot}. The XANES spectrum thus suggests that the drift is correlated with a reduction of Ge in tetrahedral sites.
It is also known from atomistic simulations based on Density Functional Theory (DFT) that tetrahedral configurations of Ge are favored by homopolar Ge-Ge bonds in GST and GeTe \cite {caravati2007coexistence,deringer2014bonding,mazza}.

DFT calculations provided some insights on the effect of aging on the electronic properties of GeTe  \cite{Raty2015,Gabardi2015,Zipoli2016}  (see Refs. \cite{MaReview,RatyReview} for a review).
Using a computational alchemy method, it was shown in Ref. \cite{Raty2015}  that the removal of Ge in tetrahedral configurations and of Ge-Ge bonds leads to a widening of the band gap and the disappearance of in-gap localized states with at the same time a lowering of the total energy. It was also shown that the disappearance of the in-gap states results from the removal of Ge-Ge bonds, whereas the widening of the band gap was resulting from  the enhancement of the Peierls distortion in pyramidal and defective octahedral configurations \cite{Raty2015}. 
At the same time, a direct simulation of the aging process was reported in our previous work \cite{Gabardi2015} by making use of a machine learning potential for large scale simulations and of the metadynamics technique to accelerate  structural transformations. The  simulations revealed that the removal of  Ge-Ge homopolar bonds and  of tetrahedra leads indeed to the reduction of localized in-gap states, to a widening of the band gap and to an overall gain in the total energy. In another study, Zipoli  et al.  \cite{Zipoli2016}  generated several small amorphous models (216 atoms) of GeTe by using a classical potential and different strategies to enhance the sampling of the potential landscape in the amorphous phase. Then, DFT calculations on these models provided  a correlation between the local structure and the optical conductivity. Once more, the change in the conductivity was ascribed to the reduction in the fraction of tetrahedral Ge and of homopolar Ge-Ge bonds, which led to a decrease in the number of in-gap states and to a shift of the Fermi closer to midgap.
 However, we mention that a combined measurement of electrical resistivity and grazing x-ray absorption spectroscopy \cite{noe-drift}  predicted an increase in the fraction of Ge-Ge  bonds with aging which is in conflict with the theoretical and EXAFS data quoted above. The reason for this discrepancy remains to be addressed.

From the theoretical side, the nature of the in-gap states in  GST was  addressed in an early DFT work for a 270-atom model \cite{caravati2009first}, where localized states at the CB edge  were assigned to chains of Sb-Te bonds, while both a midgap state and a state at the VB edge  were  localized on an  overcoodinated Ge atom in a cluster of four-membered rings reminiscent of a crystal-like environment. A subsequent analysis on four 459-atom models also revealed  states at the CB tail localized on tetrahedral Ge atoms and states at the VB tail localized on p-type lone pairs of Te \cite{caravatiblyp}, but a comprehensive statistical analysis was not provided. 
In a later series of studies \cite{Elliott2019, Kostantinou2022,Kostantinou2023}   a machine learning  Gaussian Approximation Potential (GAP) \cite{mocanu2018modeling}  was exploited to generate several amorphous models (thirty 315-atom models and one 900-atom model) by quenching from the melt.  The mid-gap states of these models obtained by DFT calculations were localized mainly on over-coordinated (5- and 6-fold)  Ge atoms in a crystal-like local environment \cite{Elliott2019}.  Other states at CB edge  were shown to act as traps for electrons that become localized  on chains of axial bonds (bonding angles close to 180$^o$ in defective octahedral geometries) \cite{Kostantinou2022}. However, this analysis was based on amorphous models featuring a strong underestimation of the fraction of homopolar Ge-Ge bonds and tetrahedral geometries compared to other DFT models in literature \cite{elliot,caravati2007coexistence,akola2007structural,Massobrio2023}. As these latter structural features were shown to control the in-gap states in GeTe, it is still uncertain whether they may play an important role in GST as well.

To address this issue, in this work we exploit a different machine learning interatomic potential (MLIP) for GST that we have recently devised and which better reproduces the fraction of wrong bonds in the amorphous phase (homopolar and Ge-Sb bonds that are not present in the crystalline phases) \cite{omar2024,acharya2025}. Analogously to the previous work in Ref. \cite{Elliott2019}, we generated many amorphous models by quenching from the melt whose electronic structure was then analyzed within DFT by using a hybrid functional \cite{hse06},  to better reproduce the band gap. Then, metadynamics simulations \cite{laio,laiogervasio,barducci} have been performed to remove the in-gap states and mimic the drift process.

\section{Methods}
We generated several amorphous models by quenching  the melt in molecular dynamics (MD) simulations by using both a MLIP and a DFT framework. The MLIP for GST  was developed in our previous work \cite{omar2024}  using the neural network (NN) scheme implemented in the DeePMD package \cite{DeePMD4,DeePMD2,DeePMD3}. The NN was trained on  a DFT database of energies and forces of about 180,000 configurations of  small supercells (57-108 atoms), computed by employing the Perdew-Burke-Ernzerhof (PBE) exchange and correlation functional \cite{PBE}.  The  potential was validated on the structural and dynamical properties of the liquid, amorphous and crystalline phases \cite{omar2024,marcorini} and it was exploited to study the crystallization kinetics in the bulk, in heterostructures \cite{acharya2025} and in a multimillion-atom model of a PCM cell \cite{Omar2025}. NN-MD simulations were performed with the DeePMD code using the Lammps code as MD driver \cite{LAMMPS}, a time step of 2 fs, and  a Nos\'e-Hoover thermostat in the NVT ensemble \cite{noseart,hoover}. 
DFT-MD simulations were also performed using the CP2K suite of programs \cite{quickstep1,quickstep2} and the same theoretical framework used to generate the MLIP, i.e. the PBE functional and norm-conserving  pseudopotentials \cite{GTH2,GTH1}  with four, five and six valence electrons  for Ge, Sb and Te. The Kohn-Sham (KS) orbitals were expanded in a Triple-Zeta-Valence plus Polarization Gaussian-type basis set, while the charge density has been expanded in a plane-wave basis set with a cut-off of 100~Ry to efficiently solve the Poisson equation within periodic boundary conditions using the Quickstep
scheme \cite{quickstep1,quickstep2}. Brillouin Zone integration was restricted to the supercell $\Gamma$ point.

Cubic supercells of different sizes were used at the experimental density of the amorphous phase ($\rho$=0.0309 atom/{\AA}$^3$)\cite{ghezzi}. The amorphous GST (a-GST) models were generated by quenching from the melt which was prepared in turn from a random model initially heated at 2000 K for 10 ps and then equilibrated at 1000 K for 15 ps. The system was then quenched to 300 K in 100 ps and then equilibrated at this temperature  for 10 ps. In this way, we generated 30 NN 297-atom models.  Other 10 DFT 216-atom models were generated by quenching independent configurations from the same trajectory of the liquid at 1000 K. A single larger 999-atom model was also generated by NN-MD. Other four 459-atom amorphous models at a slightly different density (0.030 atoms/\AA$^3$) were taken from Ref. \cite{caravatiblyp}.
Then, the geometry of all models was optimized at 0 K (both DFT and NN models)  at the PBE level, up to a force tolerance of 10$^{-4}$ Ry/bohr. Finally, the  models were further relaxed with the hybrid functional HSE06 \cite{hse06},  to better reproduce the band gap, up to a force tolerance of 3$\cdot$10$^{-4}$ Ry/bohr. 
The electronic density of states (DOS) was computed at the HSE06 level from
KS orbitals at the supercell $\Gamma$-point broadened by a Gaussian function with a standard deviation of {13.6} meV. To quantify the localization properties of individual KS states, we computed the inverse participation ratio (IPR)
which is defined for the $i$th state by $\sum_j c_{ij}^4 / \bigl(\sum_j c_{ij}^2\bigr)^2$ where $j$ runs over the Gaussian-type orbitals (GTOs) of the
basis set, while $c_{ij}$ are the expansion coefficients of the $i$-th
KS state in GTOs. According to this definition, the value of the $IPR$
varies from $1/N$ for a completely delocalized KS state where $N$ is the total number of the Gaussian-type orbitals of the basis set, to one for
a mode completely localized on a single orbital.

\noindent
\section{Results and Discussion}
\subsection{Validation and structure of the amorphous models}
\begin{figure}[]
\centering{\includegraphics[width=0.8\textwidth,keepaspectratio]{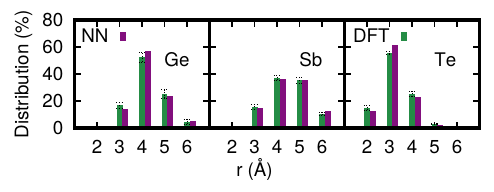}}
\caption{Distribution of the coordination numbers resolved per chemical species in the 7992-atom  a-GST model of Ref. \protect\cite{omar2024}. The distributions are computed by integrating the partial pair correlation functions up to the bonding cutoff given in Table \protect\ref{coord} (see Ref.\protect\cite{omar2024}).}
\label{coord-number}
\end{figure}

Before discussing the electronic structure of the a-GST models, we recall the structural properties of bulk amorphous GST as emerged from  previous DFT works \cite{caravati2007coexistence,akola2007structural,elliot} and from our previous work on the development of the MLIP \cite{omar2024}.
Te atoms are mostly 3-coordinated in a pyramidal geometry (three bonding angles of 90$^{\circ}$), Sb atoms are both 3-coordinated in a pyramidal geometry  and 4- or 5-coordinated in a defective octahedral environment (octahedral bonding angles but coordination lower than six),  Ge atoms are mostly in pyramidal or defective octahedral geometry with a minority fraction in tetrahedral geometries favored by Ge-Ge and Ge-Sb bonds \cite{caravati2007coexistence}. Defective octahedral geometries involve axial bonds, i.e. a pair of bonds with bonding angle close to 180$^o$. A quantitative measure of the fraction of tetrahedral environments in the NN models of a-GST was obtained in Ref. \cite{omar2024} from the local order parameter $q$ introduced in Ref.~\cite{qparam}. It is defined as $q=1-\frac{3}{8}\sum_{\rm i > k} (\frac{1}{3} + \cos \theta_{ \rm ijk})^2$,   where the sum runs over the pairs of atoms bonded to a central atom $j$ and forming a bonding angle $\theta_{\rm ijk}$. The order parameter is $q$=1 for the ideal tetrahedral geometry of 4-coordinated atoms, to $q$=0 for the 6-coordinated octahedral site, to $q$=5/8 for a 4-coordinated defective octahedral site, and $q$=7/8 for a pyramidal geometry. A fraction of Ge atoms in a tetrahedral geometry (Ge$_{\rm T}$) of about 28\% was obtained in Ref. \cite{omar2024} by integrating the $q$ parameter between 0.8 and 1 as discussed in  previous works \cite{spreafico}. 
The large majority of bonds are Ge-Te and Sb-Te bonds with a small fraction of wrong bonds (homopolar Ge-Ge, Sb-Sb, Te-Te bonds and Ge-Sb bonds) as shown in Table \ref{wrong_bonds} for a {7992}-atom model generated by NN-MD in Ref. \cite{omar2024}. The partial pair coordination numbers for the same model given in Table \ref{coord} shows a very good agreement between NN and DFT results for the wrong bond as already anticipated.
We also report in Fig. \ref{coord-number} the distribution of the coordination numbers for the different species  in the 7992-atom  a-GST model of Ref. \cite{omar2024} that will be useful in the analysis of the in-gap states.

\begin{table}[]
\begin{center}
\begin{tabular}{ccccc}
\hline 
&  & with Ge & with Sb & with Te \\
\hline
&Ge &2.34  &3.50   &42.57 \\
&Sb &    - &3.38   &43.71 \\
&Te &    - &    -  &4.48 \\
\hline
\end{tabular}
\caption{ Fraction ($\%$) of the different types of bonds over the total number of bonds in the 7992-atom model of a-GST from NN simulations of Ref. \cite{omar2024}.} \label{wrong_bonds}
\end{center}
\end{table}

\begin{table*}
\begin{center}
\begin{tabular}{ c c c c c c  }
\hline
        & Ge                   & Sb                 & Te \\ \hline
Total   & 4.19 (4.17$\pm$0.10)  &4.45 (4.40$\pm$0.05)&3.14 (3.16$\pm$0.05)\\
With Ge & 0.38 (0.34$\pm$0.05) &0.29 (0.26$\pm$0.05)&1.41 (1.42$\pm$0.02)\\
With Sb & 0.29 (0.26$\pm$0.05) &0.55 (0.51$\pm$0.10) &1.44 (1.45$\pm$0.05)\\
With Te & 3.52 (3.57$\pm$0.05) &3.61 (3.63$\pm$0.10) &0.29 (0.28$\pm$0.02)\\
\hline
\\
\end{tabular}
\caption{Average coordination numbers for different pairs of atoms in a-GST at 300 K generated from  DFT (216-atom models, in parenthesis) and NN ({7992}-atom model) simulations  from Ref. \cite{omar2024}. Error bars for DFT data are obtained from the analysis of four models. The bonding cutoff are those chosen in Ref. \cite{omar2024} corresponding to 3.2 {\AA } for all pairs except for Sb-Te for which a longer cutoff value of 3.4 {\AA } was used.}
\label{coord}
\end{center}
\end{table*}

\subsection{Electronic structure and in-gap states}
\begin{figure}[]
\centering{\includegraphics[width=0.8\textwidth,keepaspectratio]{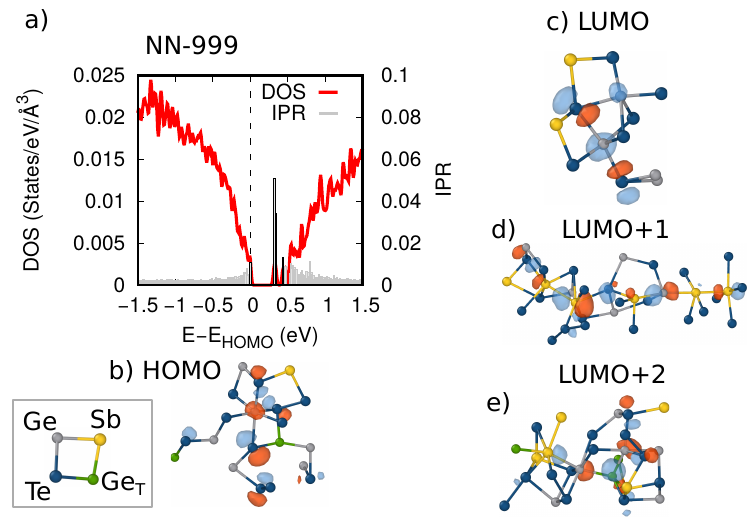}}
\caption{a) Electronic density of states (DOS) and Inverse Participation Ratio (IPR) of the 999-atom  model. The zero of energy is the highest occupied state (highest occupied molecular orbital, HOMO). b)-e) Isosurface of the most localized KS states, the HOMO (hole trap), the LUMO and LUMO+1 deep electron traps, and the LUMO+2 shallow electron trap. Ge, tetrahedral Ge (Ge$_{\rm T}$), Sb and Te atoms are depicted with gray, green, yellow and blue spheres (see color code in panel b). Isosurfaces with different signs are depicted with different (orange or azure) colors. The LUMO is localized on the axial Ge-Te bonds of an overcoordinated Ge atom. The LUMO+1 is localized on the axial bonds of a long chain Sb-Sb-Ge-Sb-Te-Sb-Sb with wrong bonds. The LUMO+2 is localized on a Sb-Ge-Ge-Ge chain with wrong bonds atoms, which involves overcoordinated Sb/Ge and tetrahedral Ge atoms..}
\label{DOS-999}
\end{figure}

Turning now to the discussion of the electronic properties, 
we first discuss the electronic structure of the large 999-atom model after which we will analyze  the nature of the in-gap states emerged from the  statistics on the small models.
The DOS and IPR of the 999-atom model is shown in Fig. \ref{DOS-999} which also display the isosurface of the most localized KS state: the highest occupied molecular orbital (HOMO) which is a hole trap, the lowest occupied molecular orbital (LUMO) and the LUMO+1 which are deep electron traps, and
the LUMO+2  shallow electron trap.
The HOMO is localized on a 6-coordinated Ge atom in a crystal-like configuration. The LUMO is localized on the  axial Ge-Te bonds of an overcoordinated Ge atom. The LUMO+1 is localized on the axial bonds of a long chain  Sb-Sb-Ge-Sb-Te-Sb-Sb with wrong bonds. The LUMO+2 is localized on a Sb-Ge-Ge-Ge chain with wrong bonds, which involves overcoordinated Sb/Ge and tetrahedral Ge atoms. Therefore, although we confirm the presence of in-gap states due to overcoordinated Ge atoms and axial bonds as found in Ref. \cite{Elliott2019}, wrong bonds and tetrahedra are also involved in the localized states.

\begin{figure}[]
\centering{\includegraphics[width=0.8\columnwidth, keepaspectratio]{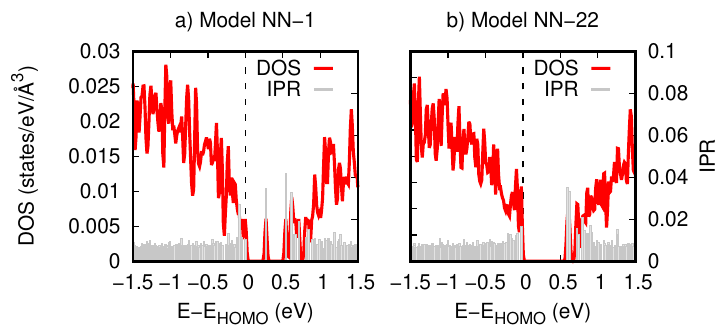}}
\caption{Electronic density of states (DOS) and Inverse Participation Ratio (IPR) of two 297-atom NN models a) with and b) without in-gap states. The zero of energy is the HOMO.}
\label{DOS-297}
\end{figure}

The analysis of a single model is obviously not sufficient to draw reliable conclusions on the nature of the in-gap states. We then  move now to the analysis of the 40 small models from which we could extract reliable statistics. As an example,  in Fig. \ref{DOS-297} we show the DOS and IPR of two 297-atom NN models with (panel a) and without (panel b) in-gap states. The corresponding plots for all the other models are shown in Figs. S1-S44 in the Supplementary Material. For each model, we assigned a mobility gap by identifying the VB edge as the energy below which all states have a  IPR lower than 0.018 and are therefore delocalized and similarly for the CB edges. Due the small size of our simulation cells, this procedure is subject to some arbitrariness and uncertainties. The resulting average gap and mean root deviation is 0.75 $\pm$ 0.085 eV which is good agreement with previous DFT calculations with hybrid functionals \cite{caravati2009first,Elliott2019} and with the experimental value of 0.7 eV \cite{Abelson}.  The mobility gap of all 40 models is given in Table S1 in the Supplementary Material. However, we must consider that the addition of the spin-orbit coupling (SOC), here neglected, substantially reduces  the band gap in GST, for example, in the hexagonal phase \cite{Tian2023}. The agreement with experiments of the HSE band gap for a-GST  by neglecting SOC might therefore be fortuitous. Anyway, for the purpose of identifying the in-gap states, a correct band gap, even though fortuitous, is sufficient here. The in-gap empty states that are potential electron traps (acceptor) have been classified as shallow  when they are within 0.2 eV from the CB  edge or 0.2 from the VB edge  and deep  otherwise. In addition a few filled localized states appear below the HOMO and above at the VB edge (mobility edge). These states are potential hole traps. Due to the small size of our simulation cells, we cannot distinguish between real shallow states separated by an energy gap from the continuous Urbach tails and the localized states in the Urbach tails themselves. Therefore, for our purpose, all states that are not deep are generically classified as shallow, either empty or filled. The shallow empty states close to the CB represent the CB tail, while the filled states  close to the VB represent the VB tail.

A fraction of 23\% of the NN models display a clean gap without deep states. This percentage increases to 50\% in DFT models. 
The average energy of the NN (DFT) models with or without deep in-gap states is 180.170 $\pm$ 0.005 (180.172 $\pm$  0.002) eV/atom or 180.171 $\pm$ 0.002 (180.171 $\pm$ 0.007) eV/atom. The difference in the average energy between the  models with and without deep in-gap states is therefore  smaller than the spread in energy of the models within the same class.

\begin{figure}[]
    \centering{
    \includegraphics[width=0.8\columnwidth, keepaspectratio]{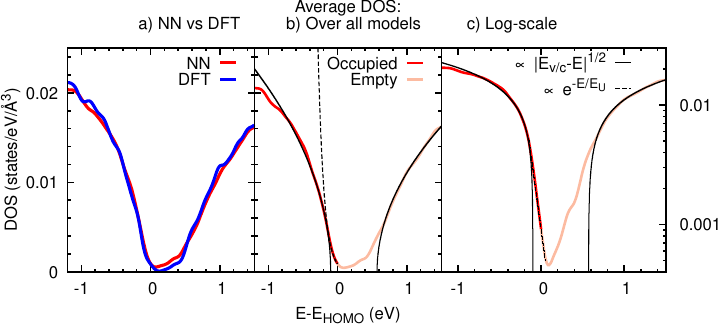}}
    \caption{a) Comparison of the electronic DOS averaged over NN models (red curve) and over DFT models (blue curve) close to the band gap. All the DOS are aligned to the bottom of the deep 5s band of Te. The zero of energy is the HOMO assigned by the integration of the average DOS. b) Electronic DOS averaged over all models. The DOS at the VB and CB edges is fitted with a band-like contribution (black continuous curves) $N_v\sqrt{E_v-E}$  and $N_c\sqrt{E_c-E}$ where $E_v$=-0.099 eV and $E_c$=0.571 eV are the band edges that yield a band gap of 0.67 eV. In turn $N_{v/c}=\frac{1}{2\pi^2} (\frac{2m_{v/c}^*}{\hbar^2})^{\frac{3}{2}}$  assign the effective masses $m_v^*$= 2.16  and  $m_c^*$ = 1.8 (in electronic mass) from  $N_{v}=$ 0.0219 and $N_{v}=$ 0.0168  states/eV$^\frac{3}{2}$/\AA$^3$. The edge of the valence band above -0.099 eV is fitted by
    an Urbach tail $exp({-E}/{E_U})$ with $E_U$=85 meV.  The presence of shallow acceptor (empty) states close to the CB prevents the fitting of the Urbach tail at the conduction edge. c) The same of panel b) in semilogarithmic scale to highlight the  exponential Urbach tail at the VB edge.}
    \label{fig:avgdos}
\end{figure}

The DOS averaged over all NN models  displays in-gap states in the energy range 0.1-0.3 eV (HOMO at zero energy) which is clean in the DOS averaged over the DFT models  (Fig. \ref{fig:avgdos}).
This discrepancy can be partially ascribed to the different sizes of the DFT (216 atoms) and NN (297 atoms) models. In fact, no in-gap states in the energy range were found in other four NN 216-atom models with the same size of the DFT ones (see Fig. S45 in the Supplementary Material). 
 We also checked that the states in the energy range 0.1-0.3 eV are not an artifact of the NN potential as follows. For a couple of different NN models we either performed a short (6 ps) DFT annealing at 300 K, or we started from the configuration in the NN simulations at 600 K from which the model was generated, and we continued the quenching protocol by DFT simulations. In all cases the resulting models (relaxed at the HSE level)  showed in-gap states in  0.1-0.3 eV range, which seems to exclude issues with the transferability of NN potential (see Figs. S46-S47 in the Supplementary Material).
 
  The DOS averaged over all models below (above) the VB (CB) edge, where KS states are delocalized,  is fitted  by the function  $N_{v/c}\sqrt{E_{v/c}-E}$ (see Fig. \ref{fig:avgdos}b) where $E_{v/c}$ is the top of the VB/CB and $N_{v/c}$ is  given by $\frac{1}{2\pi^2} (\frac{2m_{v/c}^*}{\hbar^2})^{\frac{3}{2}}$ with effective mass $m_{v/c}^*$. 
The resulting fitting parameters are given in the caption of Fig. \ref{fig:avgdos} which yield a band gap  $E_{c}-E_{v}$=0.67 eV  slightly smaller than the value estimated above (0.75 $\pm$ 0.085 eV) from the analysis of the IPR.
On the other hand, localized states in the Urbach tail of the VB are fitted  with $exp(-E/{E_U})$  and an Urbach energy $E_U$=85 meV
(see Fig. \ref{fig:avgdos}b-c). 
We will comment on this number in comparison with experimental data later on.
The  CB tail cannot be fitted by an Urbach function probably because of the presence of a band of shallow donor states superimposed to the exponential tail (see Fig. \ref{fig:avgdos}c). 

Out of the 168 in-gap states of our 40 models, 22 are filled  states close the VB maximum (lower mobility edge) which correspond to a density of 0.62 10$^{20}$/cm$^3$ (22 $\cdot$ 0.030 \AA$^{-3}$/(216 $\cdot$ 10 + 297 $\cdot$ 30)), 8 are empty shallow states close to the VB (0.22 10$^{20}$/cm$^3$), 74 are empty shallow states close to the CB (2.08 10$^{20}$/cm$^3$), and 64 are deep states (1.80 10$^{20}$/cm$^3$). 
This analysis suggests that the filled Urbach tail at the valence band is narrower than the empty Urbach tail at the CB edge.
Experimentally, in-gap states have been inferred from photoconductivity measurements \cite{Kaes2016}, consisting of  a Gaussian distribution of deep acceptor (i.e. empty) states  0.1 eV below midgap and a Gaussian distribution of shallow donor (i.e. filled) states 0.2 eV below midgap and close to the VB (see Fig. 1 in Ref.\cite{Kaes2016}). Each distribution corresponds to a defects density of about 10$^{20}$/cm$^3$. The density of deep acceptor states is consistent with our results (see above). On the other hand, due to the small size of our simulation cells, we have not been able to disentangle a possible distribution of shallow filled states (donors) close to the VB  from the VB tail itself described by an Urbach function with $E_U$=85 meV. 
Experimentally, in-gap states have been inferred from photoconductivity measurements \cite{Kaes2016}, consisting of  a Gaussian distribution of deep acceptor (i.e. empty) states  0.1 eV below midgap and a Gaussian distribution of shallow donor (i.e. filled) states 0.2 eV below midgap and close to the VB (see Fig. 1 in Ref.\cite{Kaes2016}). Each distribution corresponds to a defects density of about 10$^{20}$/cm$^3$. The density of deep acceptor states is consistent with our results (see above). On the other hand, due to the small size of our simulation cells, we have not been able to disentangle a possible distribution of shallow filled states (donors) close to the VB  from the VB tail itself described by an Urbach function with $E_U$=85 meV. 
Experimentally, an Urbach energy of 81 meV \cite{Abelson} is measured from the optical absorption spectra which is ruled by the joint density of valence and conduction states (JDOS). The JDOS is proportional to $(exp(E/E_{U_v}) - exp(E/E_{U_c}))/(E_{U_v}-E_{U_c})$, where $E_{U_v}$ and $E_{U_c}$ are the Urbach energies of valence and conduction bands. For a band gap much larger than the Urbach energies, the measured  $E_U$ corresponds to the largest between $E_{U_v}$ and $E_{U_c}$.
According to the considerations given above on the number of localized states close to the VB and CB, we thus expect that the measured  $E_U$ should correspond to the conduction Urbach energy which, however, cannot be extracted from our data due to the superposition of a band of shallow states to the exponential tail.

 \subsection{Local environment of in-gap states}

We analyzed the atomic environment on which each in-gap state is localized by inspecting the isosurface of the KS orbitals shown in Figs. S1-S44 in the Supplementary Material. The isosurface is drawn for  a value of the orbital  of 50-70 $\%$ of the maximum value.  This analysis was performed for 80 in-gap states, i.e. about one half of the total number of states distributed as 6 filled states close the VB, 2 empty shallow states close to the VB, 32 empty shallow states close to the CB and 40 deep states. To these latter 80 states, we added additional four filled states in the CB tail from the 459-atom models to improve the statistics on the localized filled states which are relatively few in all our models.
We selected these states because of their highest IPR in each model. The states selected for this analysis are highlighted in the DOS with bold IPR for all models in Figs. S1-S44 in the Supplementary Material.

\begin{figure}[]
    \centering{
    \includegraphics[width=0.8\columnwidth, keepaspectratio]{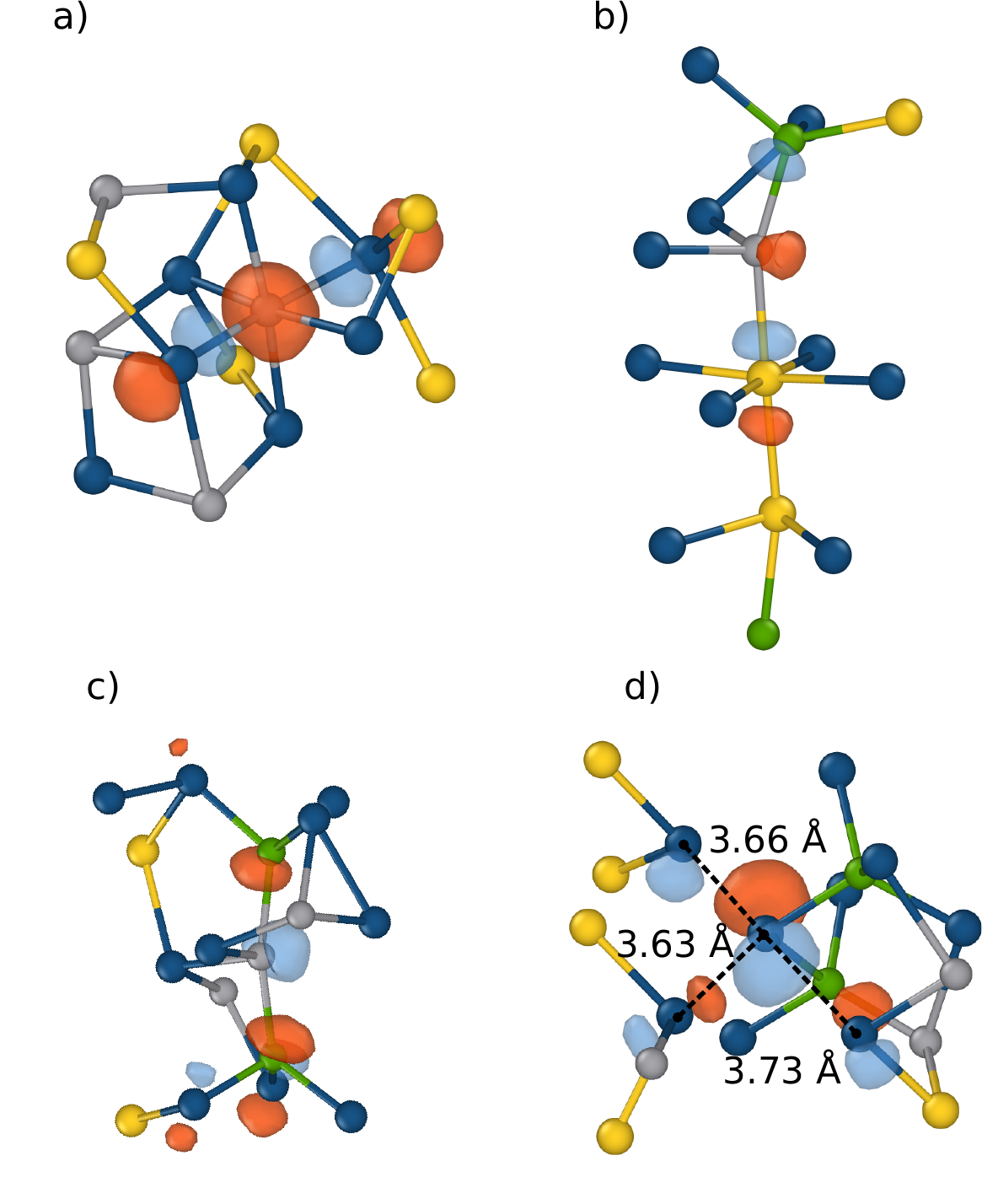}}
    \caption{Example of localized in-gap states showing the typical features of the atomic environment  discussed in the  text. a) An overcoordinated Ge atom in a crystal-like environment. b) A long chain of wrong bonds including  an overcoordinated Sb atom and a tetrahedral Ge atom. c) A four-coordinated Ge atom bonded to two tetrahedral Ge atoms with two wrong bonds. The central Ge atom forms two axial bonds with the two tetrahedral Ge atoms. d) A two-coordinated Te atom forming three long Te-Te contacts. The color code is the same used in Fig. \ref{DOS-999}. Ge, tetrahedral Ge (Ge$_{\rm T}$), Sb and Te atoms are depicted with gray, green, yellow and blue spheres.}
    \label{fig:features}
\end{figure}

We identified a few features of the atomic environment in which the in-gap state is localized (see Fig. \ref{fig:features}):

1) an overcoordinated, i.e. 5- or 6-coordinated Ge atom (Ge$_{over}$), 2)  an overcoordinated, i.e. 6-coordinated Sb atom (Sb$_{over}$), 3) a tetrahedral Ge atom (Ge$_{\rm T}$), 4) an atom with a wrong bond (WB), i.e Ge-Ge, Sb-Sb, Ge-Sb, Te-Te bonds. The 6-coordinated Ge and Sb atoms correspond to  
an octahedral  configuration reminiscent of the cubic rocksalt phase, while  5-coordinated Ge atoms correspond to a defective octahedra but also to a trigonal bypiramid configuration, which we also see in our models. 
Although 4-coordinated Ge and 4- and 5-coordinated Sb atoms can also be in a defective octahedral configuration with axial bonds, we do not include these configurations as a feature to classify in-gap states because of their high abundance in a-GST (see Fig. \ref{coord-number}). In other words, we attempt to classify the in-gap states with features that are present in a low fraction in the amorphous models.
We will comment later on the presence of axial bonds in localized states.
The atoms  included in this analysis are those on which a large portion of the isosurface of the KS state is localized.
It must also be stressed that the presence of these features is not a sufficient condition for an environment to be the host of a localized state.
The in-gap states are typically localized in a complex containing several atoms, and
the localization around an atom does not depend only on the nature of its first coordination shell, but also on the positions  and coordination of atoms further apart, as shown, for instance, for amorphous GeSe in Ref. \cite{InfoMatGeSe}. 
Nevertheless, we now attempt to classify the in-gap states according to the presence of the four features listed above.
The picture emerging from this analysis is summarized in Tables \ref{TableFeatures}-\ref{bonds}.
Table \ref{TableFeatures} shows the fraction of in-gap states that display the four features listed above, e.g.  the fraction of in-gap states with an overcoordinated Ge/Sb atom. This analysis is resolved in the different columns in Table \ref{TableFeatures} for the different classes of in-gap states: deep states,  CB tail (empty),  VB tail (filled). The only two empty shallow states close to the VB are included in the latter group (VB tail). The sum of each column does not yield 100 $\%$ because the large majority of in-gap states display more than one of the four features listed above. Ge$_{over}$/Sb$_{over}$, Ge$_{\rm T}$ and WB are present in a very large fraction of in-gap states. In particular, WB are present in the overwhelming majority of states at mid-gap (deep) and in CB tail, while overcoordinated Ge/Sb atoms are present in a large fraction of deep states and to a lesser extent, equal to tetrahedral Ge, in the CB tail. 
The breakdown of these distributions is shown in Table \ref{TableFeatures2}  where we report the fraction of in-gap states with only one of the features listed above, the different possible pairs of two features and the fraction of states with more than two features. The distributions are resolved for deep states,  CB tail (empty),  VB tail + shallow acceptor close to VB as in Table  \ref{TableFeatures}. The   fractions in each column is Table \ref{TableFeatures2} now sum to 100 $\%$ whereas a new feature consisting of Te-Te long contact ($>$ 3.2 \AA) is added.
Long Te-Te contacts and Te-Te wrong bonds are  present more in the VB tail than in deep and CB tail states.

\begin{table}
\begin{center}
\resizebox{\textwidth}{!}{\begin{tabular}{lccc}
\hline
    Feature  &     Deep   & VB tail + shallow acceptor  & CB tail \\
    &  (40 states) & (12 states) & (32 states) \\
    \hline
Overcoordinated Ge (Ge$_{over}$) &45.0  &41.5 &18.75    \\
Overcoordinated Sb (Sb$_{over}$) &17.5  &8.3 &12.50     \\
Tetrahedral Ge  (Ge$_{\rm T}$)         &40.0  &25.0 &56.25    \\
Wrong bonds (WB)                 &77.5  &33.2 &81.25    \\
Ge$_{over}$/Sb$_{over}$          &57.5  &41.5 &28.10    \\
More WB                          &52.5  &16.2 &50.00    \\
\hline
\\
\end{tabular}}
\caption{Fraction ($\%$) of the in-gap states in the  different classes (deep, VB tail + shallow acceptor close to the VB, CB tail) that are localized on atoms with the features listed in the first column, i.e. an overcoordinated Ge atom (5- or 6-coordinated), an overcoodinated Sb atom (6-coordinated), a tetrahedral Ge atom, an atom with at least one wrong bond (WB, see text).
 In the last two rows we also report the fraction of in-gap states with an overcoordinated cation, either Ge or Sb or both, and the fraction of in-gap with atoms with more than one wrong bond. The sum on each column does not yield 100 $\%$ because the large majority on in-gap states display more than one of the four features listed above.}
\label{TableFeatures}
\end{center}
\end{table}

\begin{table}
\begin{center}
\resizebox{\textwidth}{!}{\begin{tabular}{lccc}
\hline
    Feature  &     Deep   & VB tail + shallow acceptor  & CB tail \\

    &  (40 states) & (12 states) & (32 states) \\
    \hline
Only Ge$_{over}$/Sb$_{over}$     &12.5 & 8.3 &  3.125 \\
Only Ge$_{\rm T}$                      & 2.5 & 8.3 &  9.375 \\
Only WB                          & 7.5 & 8.3 & 21.875 \\
Ge$_{over}$/Sb$_{over}$ + WB     &30.0 & 16.6 &  9.375 \\
Ge$_{\rm T}$ + WB                      &25.0 &  8.3 & 34.375 \\
Ge$_{over}$/Sb$_{over}$ + Ge$_{\rm T}$ &-    & -    &  -     \\
more than 2 features             &15.0 & 16.6 & 15.625 \\
Only Te-Te long contact          &5    & 33.2 &  - \\
\hline
\\
\end{tabular}}
\caption{Fraction ($\%$) of the in-gap states in the  different classes (deep, VB tail + shallow acceptor close to the VB, CB tail) that display one or more of the different four features (see text and Table \ref{TableFeatures}).}
\label{TableFeatures2}
\end{center}
\end{table}

The vast majority of in-gap states display more than one of the four features listed above. 
To gain insight into the role that the different features could have in the localization of KS orbitals, we analyzed whether these features are overrepresented in the in-gap states with respect to the bulk. This  analysis is summarized in Table \ref{TableDef-vs-Bulk} which reports the fraction of atoms with a particular feature (Ge$_{over}$/Sb$_{over}$, Ge$_{\rm T}$, atom with a WB), among those defined to belong to the environment of the in-gap states, compared to the corresponding fraction among all atoms in the bulk of the 7992-atom amorphous model.

\begin{table}
\begin{center}
\begin{tabular}{ccc}
\hline
Feature  &  in-gap states  & bulk \\
\hline
Ge$_{over}$/Sb$_{over}$  &26 & 22\\
Ge$_{\rm T}$  & 49& 25\\
WB &50 & 35\\
\hline
\end{tabular}
\caption{Fraction ($\%$) of atoms that host the in-gap states (see text) that are either an overcoordinated Ge or Sb atom (Ge$_{over}$ or Sb$_{over}$), a tetrahedral Ge atom (Ge$_{\rm T}$), or an atom with a wrong bond (WB), compared with the corresponding fraction in the bulk (7992-atom model of Ref. \cite{omar2024}).}
\label{TableDef-vs-Bulk}
\end{center}
\end{table}

\begin{table}
\begin{center}
\begin{tabular}{ccc}
\hline
Bonds  &  in-gap states  & bulk \\
\hline
Ge-Ge  &17 & 16\\
Ge-Sb&32& 25\\
Sb-Sb&23 & 24\\
Te-Te& 28 & 35\\
\hline
\end{tabular}
\caption{Fraction ($\%$) of the different types or wrong bonds among those present in the in-gap states (second column) and in the bulk (third column). The wrong bonds in the in-gap states are those involving the atoms which are defined to belong to the environment that hosts the localized state according to the isosurface of the corresponding KS orbital.}
\label{bonds}
\end{center}
\end{table}

It is clear from Table \ref{TableDef-vs-Bulk} that both Ge$_{\rm T}$ atoms and wrong bonds are overrepresented in the in-gap states with respect to the bulk. These two features are not independent, since the tetrahedral geometry of Ge is favored by Ge-Ge and Ge-Sb wrong bonds. 
This result is not surprising, since these structural motifs are not present in the crystal and  are presumably weaker structures. Indeed, wrong bonds were demonstrated to be weaker than Ge-Te and Sb-Te in liquid GST from the analysis of the integrated crystal orbital Hamilton population (ICOHP)
extracted from the DFT electronic structure \cite{lee2025wrongbonds}.
Among the wrong bonds, the Ge-Sb are the most abundant and more overrepresented in the in-gap states with respect to the bulk, as shown in Table \ref{bonds}. In GeTe, the DFT analysis on 1728-atom models of Ref. \cite{Gabardi2015} showed that most the in-gap states were localized on short chain of Ge-Ge bonds. On the contrary, in our GST models only about 46 $\%$ of the in-gap states comprise atoms with more 
than one wrong bond (see Table \ref{TableFeatures}). 
Although the environment of the overwhelming majority of the in-gap states includes  wrong bonds 
(i.e. atoms with at least one wrong bond), states with only wrong bonds and without tetrahedra or overcoordinated atoms are rare (see Table \ref{TableFeatures2}).
Wrong bonds favor tetrahedra and thus these two features mostly go together. In addition, wrong bonds also accompany overcoordinated atoms in the vast majority of in-gap states in which overcoordinated atoms are present.
Some of these configurations, i.e. overcoordinated atoms plus wrong bonds, could be seen as a sort of crystal-like environment with an antisite defect. Wrong bonds were also present in some of the crystal-like  environments that host mid-gap states in Ref. \cite{Elliott2019}. 
We finally remark that about 45 $\%$ of the atoms on which the  in-gap states are localized feature at least one axial bond, while
this fraction  in the bulk is about 35 $\%$.

\subsection{Metadynamics simulations mimicking aging}
Having identified the nature of in-gap states, we now attempt to remove them by inducing structural relaxations.
We mimicked this aging process using the metadynamic technique. This method \cite{laio,laiogervasio,barducci} is based on a coarse-grained, non-Markovian dynamics in the manifold spanned by few reaction coordinates (collective variables), biased by a history-dependent potential, which drives the system towards the lowest saddle point. The main assumption is that the reaction path could be described in the manifold of few collective coordinates $S_{\alpha}(\{{\bf R}_I\})$ which are functions of the ionic coordinates ${\bf R}_I$. A history-dependent   external potential $V(t,S_{\alpha}(\{{\bf R}_I\}))$  is applied to the ionic coordinates and constructed by the accumulation of Gaussian functions, centered at the positions of the $\{S_{\alpha}\}$ already visited along the trajectory. The potential discourages the system from remaining in the region  already visited and pushes it over the lowest energy barrier towards a new local minimum. Metadynamics has been  applied  to study several chemical reactions and structural transformation in the gas and condensed phase \cite{laiogervasio,barducci} and also to study aging in amorphous GeTe in our previous work \cite{Gabardi2015}.

\begin{figure}[]
    \centering
    \includegraphics{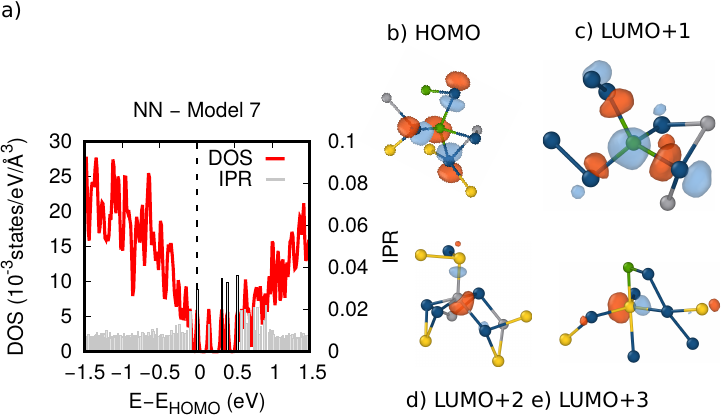}
    \caption{A zoom of the electronic DOS of model NN-7 (red curve) close to the band gap and the value of the IPR (gray spikes) of KS states at $\Gamma$ point. The zero of energy is the top of the valence band. The isosurface of states with high IPR (highlighted with black border) are shown in panels b)-e). b) The HOMO   is localized on a 4-coordinated Ge atom with a q-order parameter of 0.78 very close to that of a G$_T$ atom. c) The LUMO+1 is localized on a Ge$_{\rm T}$ atom. d) The LUMO+2 is localized on a 5-coordinated Ge atom with a wrong bond. e) LUMO+3 is localized on a 5-coordinated Sb atom bonded to a Ge$_{\rm T}$ atom. Ge, Ge$_{\rm T}$, Sb and Te are depicted with gray, green, yellow and blue spheres.}
    \label{fig:NN7}
\end{figure}

\begin{figure}[]
    \centering
    \includegraphics[width=0.9\textwidth,keepaspectratio]{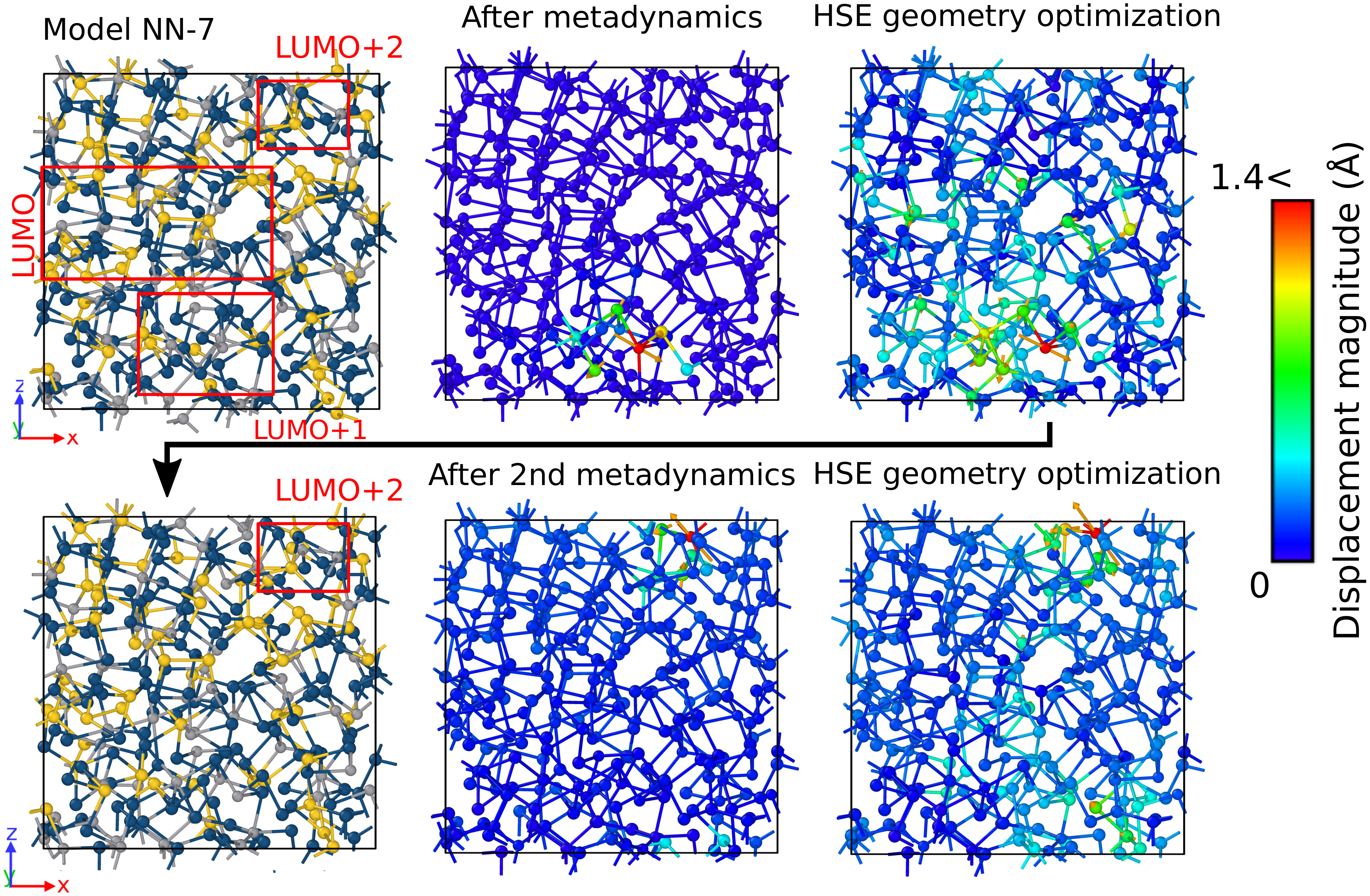}
    \caption{a) Snapshots of the metadynamics simulation of the aging process of mdel NN-7  consisting of four  steps: metadynamics by biasing only the region around the LUMO+1 state, HSE geometry optimization, metadynamics by biasing only the region of former LUMO+2 state, HSE geometry optimization. The region around the LUMO, LUMO+1 and LUMO+2 states are  highlighted in the first panel (see the text). The displacement of the atoms is highlighted by the color scale.}
    \label{fig:meta-7}
\end{figure}

\begin{figure}[]
    \centering
    \includegraphics[width=0.9\textwidth,keepaspectratio]{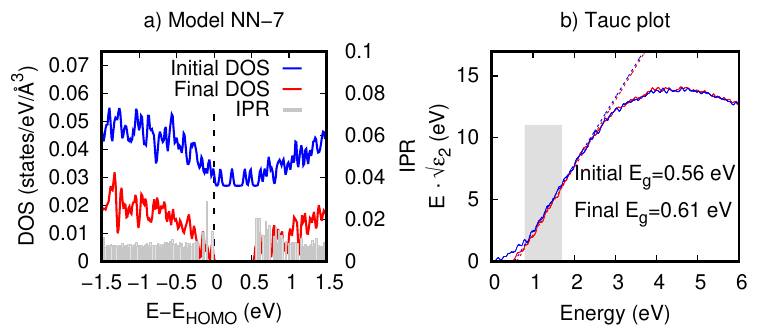}
    \caption{a) Electronic DOS  (shifted) of the initial configuration of model NN-7 superimposed to the DOS and IPR of the HSE-optimized configuration after the two metadynamics runs (see Fig. \protect\ref{fig:meta-7}).  b) Tauc plot and Tauc gap given as an inset of the initial configuration and the final (HSE optimized) configurations after the two metadynamics runs (see Fig. \ref{fig:meta-7}).}
    \label{fig:meta-7-DOS}
\end{figure}
Here, as collective variable we choose the potential energy itself \cite{bonomi} which sounds a reasonable variable to describe the relaxation of the glass.  We performed metadynamics simulations on the model NN-7 
whose electronic structure and character of the localized states are shown in Fig. \ref{fig:NN7}.
 On the assumption that aging should lead to a removal of the in-gap,  we applied the biasing potential only to the regions where the in-gap states are localized with all other atoms fixed (see Fig. \ref{fig:meta-7}). The selected mobile regions where the LUMO+1 and LUMO+2 states are localized are shown in Fig. \ref{fig:meta-7}a.  We performed two metadynamics runs. We first performed NN-MD simulation by selecting only the LUMO+1  region  around a Ge$_{\rm T}$ atom, then we optimized the geometry at the HSE level. Transformation of the Ge atom from a tetrahedral geometry to a pyramidal 3-coordinated geometry leads to removal of the in-gap state (former LUMO+1) but also of the neighboring poorly localized LUMO, as shown by the resulting DOS and IPR in Fig. \ref{fig:meta-7-DOS}a. Then, we performed a second metadynamics simulation by biasing the region of the former LUMO+2 state (5-coordinated Ge atom with a wrong bond) and relaxed the final configuration at the HSE level. The former LUMO+2 but also LUMO+3 (5-coordinated Sb atom bonded to a Ge$_{\rm T}$ atom) in-gap states were removed as shown by the DOS in Fig. \ref{fig:meta-7-DOS}b. In the transformation, the 5-fold Ge atom (LUMO+2) turns into a 4-coordinated atom by breaking a Ge-Te bond, while the environment of LUMO+3 undergoes just small deformation without coordination changes. The simulations show that in-gap states can be removed in different manners, either by removing tetrahedra or reducing the coordination of overcoordinated atoms, but also by relaxing the environment without an apparent change in bonding of atoms on which the state is localized. 
The transformation leads to the cleaning of the band gap and to an overall energy gain of 0.6 eV for the removal of four in-gap states.

As a proxy of the mobility gap, we computed the Tauc gap \cite{refTauc} which is defined by the intercept with the energy axis of the linear extrapolation of  $\omega \sqrt{\varepsilon_2(\omega)}$ as a function of $\omega$, where $\varepsilon_2(\omega)$ is the imaginary part of the dielectric function.  In turn, we computed $\varepsilon_2(\omega)$ in the random phase approximation (RPA) from Kohn-Sham (KS) orbitals within DFT
\begin{equation}
\varepsilon_2(\omega) =  \frac{8 \pi^2}{3 V_o \omega^2} \sum_{v,c}  |\langle c | {\bf p} |v \rangle |^2 \delta(\omega - E_c + E_v),
\label{EQ1}
\end{equation}
where $E_c$ and $E_v$ refer to the energies of conduction and valence bands at the $\Gamma$-point of the supercell BZ. $V_o$ is the unit cell volume.
The Tauc plot and the resulting Tauc gap of the initial configuration and the final configuration after the two metadynamics runs are shown in Fig. \ref{fig:meta-7-DOS}. In addition to the removal of the in-gap states, the aging process mimicked by the metadynamics simulations also leads to a widening of the band gap by 50 meV which is similar to the band gap increase of 42 meV measured experimentally by optical absorption after 172 hours of aging at room temperature \cite{fantini_optical}, since metadynamics accelerates the  aging by biasing structural transformation.

\section{Conclusions}
In summary, we have provided an extensive statistical analysis on the origin of in-gap states in GST. We analyzed 44 models of different sizes (216-459 atoms) optimized at the DFT level and generated by quenching the melt either in DFT-MD or NN-MD. The in-gap states have been classified as deep or shallow, the latter also include  the Urbach tails. The density of deep states is of the same order of magnitude of the value inferred from photoconductivity measurements \cite{Kaes2016}. The analysis of the local environment of the in-gap states reveal the presence of three main features: wrong bonds (homopolar or Ge-Sb), Ge atoms in tetrahedral geometry, overcoordinated Ge and Sb in an octahedral geometry (5- and 6-coordinated Ge and 6-coordinated Sb which correspond to a minority fraction in the bulk). The vast majority of in-gap states display more than one of the features listed above, but overall 82 $\%$ of in-gap states are localized on atoms with at least one wrong bond. Overcoordinated atoms in an octahedral geometry, also identified in a previous work \cite{Elliott2019} in crystal-like environment, are present in 58 $\%$ of the in-gap states, but they are mainly accompanied by a wrong bond. Some of these correspond to  a sort of antisite crystal-like configurations. Tetrahedral Ge atoms are present in about 40 $\%$ of the in-gap states, but again nearly always together with a wrong bond. Therefore, we can conclude that wrong bonds play an important role in the formation of most of the in-gap states as was found in GeTe \cite{Raty2015,Gabardi2015} and more recently in the  selenide alloys such as GeSe$_x$ \cite{slassi} and GeAsSe \cite{CaravatiSSS}. However, unlike GeAsSe \cite{CaravatiSSS}, no valence alternation pairs (VAPs) are found in GST. As wrong bonds and tetrahedral Ge atoms are supposed to disappear upon aging towards a more ideal glass, our findings support the picture that the removal of these structural features during aging would lead to an increase of resistance because of the depletion of the in-gap states that support the Poole-Frenkel conduction mechanism. These results extend to GST the picture already emerged from DFT simulations for GeTe \cite{Raty2015,Gabardi2015,Zipoli2016}. 
Metadynamics simulations  mimicking the aging of a single model show 
indeed the removal of a tetrahedron and of the corresponding in-gap state, but also other processes. Some in-gap states are removed by reducing the coordination of overcoordinated atoms, while others are removed by relaxing the environment without an apparent change in bonding of atoms on which the states are localized. 
The simulations show an energy gain of about 0.6 eV during aging, the simultaneous removal of four in-gap states and the widening of the Tauc gap by about 50 meV which is comparable to the experimentally measured shift during aging \cite{fantini_optical}. 
It is conceivable that once wrong bonds and tetrahedra were removed after aging, the surviving in-gap states would be mostly localized on overcoordinated atoms in crystal-like geometries (with no wrong bonds) as those found in Ref. \cite{Elliott2019} on amorphous DFT models closer to the ideal glass. These results thus provide important insights on the link between the evolution of the in-gap states responsible for the resistance drift and the structural relaxations of the glass, which is of great relevance for the mitigation of the drift in phase change memory devices.

\bibliographystyle{elsarticle-num} 
\bibliography{ref.bib}

\begin{thebibliography}{10}
\expandafter\ifx\csname url\endcsname\relax
  \def\url#1{\texttt{#1}}\fi
\expandafter\ifx\csname urlprefix\endcsname\relax\def\urlprefix{URL }\fi
\expandafter\ifx\csname href\endcsname\relax
  \def\href#1#2{#2} \def\path#1{#1}\fi

\bibitem{wuttig2007phase}
M.~Wuttig, N.~Yamada, Phase-change materials for rewriteable data storage, Nat. Mater. 6 (2007) 824--832.

\bibitem{noe2017phase}
P.~No{\'e}, C.~Vall{\'e}e, F.~Hippert, F.~Fillot, J.-Y. Raty, Phase-change materials for non-volatile memory devices: from technological challenges to materials science issues, Semicond. Sci. Technol. 33 (2017) 013002.

\bibitem{fantini}
P.~Fantini, Phase change memory applications: the history, the present and the future, J. Phys. D: Appl. Phys. 53 (2020) 283002.

\bibitem{Cappelletti_2020}
P.~Cappelletti, R.~Annunziata, F.~Arnaud, F.~Disegni, A.~Maurelli, P.~Zuliani, Phase change memory for automotive grade embedded {NVM} applications, J. Phys. D 53 (2020) 193002.

\bibitem{Redaelli2022}
A.~Redaelli, E.~Petroni, R.~Annunziata, {Material and process engineering challenges in Ge-rich GST for embedded PCM}, Materials Science in Semiconductor Processing 137 (2022) 106184.

\bibitem{kuzum2012nanoelectronic}
D.~Kuzum, R.~G. Jeyasingh, B.~Lee, H.-S.~P. Wong, Nanoelectronic programmable synapses based on phase change materials for brain-inspired computing, Nano Lett. 12 (2012) 2179--2186.

\bibitem{tuma2016stochastic}
T.~Tuma, A.~Pantazi, M.~Le~Gallo, A.~Sebastian, E.~Eleftheriou, Stochastic phase-change neurons, Nat. Nanotechnol. 11 (2016) 693--699.

\bibitem{sebastianNanotech}
A.~Sebastian, M.~Le~Gallo, R.~Khaddam-Aljameh, E.~Eleftheriou, Memory devices and applications for in-memory computing, Nat. Nanotechnol. 15 (2020) 529--544.

\bibitem{Boniardi2009}
M.~Boniardi, A.~Redaelli, A.~Pirovano, I.~Tortorelli, D.~Ielmini, F.~Pellizzer, A physics-based model of electrical conduction decrease with time in amorphous {G}e$_2${S}b$_2${T}e$_5$, J. Appl. Phys. 105 (2009) 084506.

\bibitem{IelminiDrift2007}
D.~Ielmini, S.~Lavizzari, D.~Sharma, A.~L. Lacaita, Physical interpretation, modeling and impact on phase change memory {(PCM)} reliability of resistance drift due to chalcogenide structural relaxation, IEEE International Electron Devices Meeting (2007) 939--942.

\bibitem{salinga2025}
J.~Ballmaier, S.~Walfort, M.~Salinga, Resistance drift of phase change materials beyond the power law, Adv. Electron. Mater. 11 (2025) 2400905.

\bibitem{multilevel}
A.~Sebastian, N.~Papandreou, A.~Pantazi, H.~Pozidis, E.~Eleftheriou, Non-resistance-based cell-state metric for phase-change memory, J. Appl. Phys. 110 (2011) 084505.

\bibitem{Papandreou}
N.~Papandreou, H.~Pozidis, T.~Mittelholzer, G.~F. Close, M.~Breitwisch, C.~Lam, E.~Eleftheriou, Drift-tolerant multilevel phase-change memory, 3rd IEEE International Memory Workshop (2011) 1--4.

\bibitem{Koelmans}
W.~W. Koelmans, A.~Sebastian, V.~P. Jonnalagadda, D.~Krebs, L.~Dellmann, E.~Eleftheriou, Projected phase-change memory devices, Nat. Commun. 6 (2015) 8181.

\bibitem{LeGalloReview}
M.~Le~Gallo, A.~Sebastian, An overview of phase-change memory device physics, J. Phys. D: Appl. Phys. 53 (2020) 213002.

\bibitem{RatyReview}
J.-Y. Raty, Aging in phase change materials: Getting insight from simulation, Phys. Status Solidi RRL 13 (2019) 1800590.

\bibitem{MaReview}
W.~Zhang, E.~Ma, Unveiling the structural origin to control resistance drift in phase-change memory materials, Mater. Today 41 (2020) 156.

\bibitem{karpov}
I.~V. Karpov, M.~Mitra, D.~Kau, G.~Spadini, Y.~A. Kryukov, V.~G. Karpov, Fundamental drift of parameters in chalcogenide phase change memory, J. Appl. Phys. 102 (2007) 124503.

\bibitem{agarwal}
M.~Mitra, Y.~Jung, D.~S. Gianola, R.~Agarwal, Extremely low drift of resistance and threshold voltage in amorphous phase change nanowire devices, Appl. Phys. Lett. 96 (2010) 222111.

\bibitem{fantini_film}
M.~Rizzi, A.~Spessot, P.~Fantini, D.~Ielmini, Role of mechanical stress in the resistance drift of {Ge$_2$Sb$_2$Te$_5$} films and phase change memories, Appl. Phys. Lett. 99 (2007) 223513.

\bibitem{Ielmini2009}
D.~Ielmini, D.~Sharma, S.~Lavizzari, A.~L. Lacaita, Reliability impact of chalcogenide-structure relaxation in {P}hase-{C}hange {M}emory ({PCM}) cells—part {I}:experimental study, IEEE Trans. Electron. Dev. 56 (2009) 1070--1077.

\bibitem{Ielmini_Poole_Frenkel}
D.~Ielmini, Y.~Zhang, Evidence for trap-limited transport in the subthreshold conduction regime of chalcogenide glasses, Appl. Phys. Lett. 90 (2007) 192102.

\bibitem{Longeaud}
C.~Longeaud, J.~Luckas, D.~Krebs, R.~Carius, J.~Klomfass, M.~Wuttig, On the density of states of germanium telluride, J. Appl. Phys. 112 (2012) 113714.

\bibitem{Luckas2013}
D.~Luckas, J.and~Krebs, S.~Grothe, J.~Klomfass, R.~Carius, C.~Longeaud, M.~Wuttig, Defects in amorphous phase-change materials, J. Mater. Res, 28 (2013) 1139--1147.

\bibitem{SalingaDrift2014}
M.~Wimmer, C.~Kaes, Matthias amd~Dellen, M.~Salinga, Role of activation energy in resistance drift of amorphous phase change materials, Front. Phys. 2 (2014) 75.

\bibitem{Lavizzari2009}
S.~Lavizzari, D.~Ielmini, D.~Sharma, , A.~L. Lacaita, Reliability impact of chalcogenide-structure relaxation in {P}hase-{C}hange {M}emory ({PCM}) cells—part {II}: Physics-based modeling, IEEE Trans. Electron. Dev. 56 (2009) 1078--1085.

\bibitem{FantiniCampoE}
P.~Fantini, M.~Ferro, A.~Calderoni, Field-accelerated structural relaxation in the amorphous state of phase change memory, Appl. Phys. Lett. 102 (2013) 253505.

\bibitem{LeGallo2018}
M.~Le~Gallo, D.~Krebs, F.~Zipoli, M.~Salinga, A.~Sebastian, Collective structural relaxation in phase-change memory devices, Adv. Electron. Mater. 4 (2018) 1700627.

\bibitem{fantini_optical}
P.~Fantini, S.~Brazzelli, E.~Cazzini, A.~Mani, Band gap widening with time induced by structural relaxation in amorphous {Ge$_2$Sb$_2$Te$_5$} films, Appl. Phys. Lett. 100 (2012) 013505.

\bibitem{Rutten2015}
M.~Rütten, M.~Kaes, A.~Albert, M.~Wuttig, M.~Salinga, Relation between bandgap and resistance drift in amorphous phase change materials, Sci. Rep. 5 (2015) 17362.

\bibitem{Wuttig2024}
J.~Pries, C.~Stenz, S.~Wei, M.~Wuttig, P.~Lucas, Structural relaxation of amorphous phase change materials at room temperature, J. Appl. Phys. 135 (2024) 135101.

\bibitem{Elliott2020}
S.~R. Elliott, Electronic mechanism for resistance drift in phase-change memory materials: link to persistent photoconductivity, J. Phys. D: Appl. Phys. 53 (2020) 214002.

\bibitem{Khan2020}
R.~S. Khan, F.~Dirisaglik, A.~Gokirmak, H.~Silva, Resistance drift in {G}e$_2${S}b$_2${T}e$_5$ phase change memory line cells at low temperatures and its response to photoexcitation, Appl. Phys. Lett. 116 (2020) 253501.

\bibitem{Yalon}
R.-G. Nir-Harwood, G.~Cohen, A.~Majumdar, R.~Haight, E.~Ber, L.~Gignac, E.~Ordan, L.~Shoham, Y.~Keller, L.~Kornblum, E.~Yalon, Drift of {S}chottky barrier height in phase change materials, ACS Nano 18 (2024) 8029--8037.

\bibitem{kolobovdrift}
K.~V. Mitrofanov, A.~V. Kolobov, P.~Fons, X.~Wang, J.~Tominaga, Y.~Tamenori, T.~Uruga, N.~Ciocchini, D.~Ielmini, {Ge L$_3$}-edge x-ray absorption near-edge structure study of structural changes accompanying conductivity drift in the amorphous phase of {Ge$_2$Sb$_2$Te$_5$}, J. Appl. Phys. 115 (2014) 173501.

\bibitem{caravati2007coexistence}
S.~Caravati, M.~Bernasconi, T.~K{\"u}hne, M.~Krack, M.~Parrinello, Coexistence of tetrahedral-and octahedral-like sites in amorphous phase change materials, Appl. Phys. Lett. 91 (2007) 171906.

\bibitem{akola2007structural}
J.~Akola, R.~Jones, Structural phase transitions on the nanoscale: The crucial pattern in the phase-change materials {G}e$_2${S}b$_2${T}e$_5$ and {G}e{T}e, Phys. Rev. B 76 (2007) 235201.

\bibitem{elliot}
J.~Heged{\"u}s, S.~Elliott, Microscopic origin of the fast crystallization ability of {G}e--{S}b--{T}e phase-change memory materials, Nat. Mater. 7 (2008) 399--405.

\bibitem{deringer2014bonding}
V.~L. Deringer, W.~Zhang, M.~Lumeij, S.~Maintz, M.~Wuttig, R.~Mazzarello, R.~Dronskowski, Bonding nature of local structural motifs in amorphous {G}e{T}e, Angew. Chem. Int. Ed. 53 (2014) 10817--10820.

\bibitem{mazza}
R.~Mazzarello, S.~Caravati, S.~Angioletti-Uberti, M.~Bernasconi, M.~Parrinello, {Signature of Tetrahedral Ge in the Raman Spectrum of Amorphous Phase-Change Materials}, Phys. Rev. Lett. 104 (2010) 085503.

\bibitem{Raty2015}
J.-Y. Raty, W.~Zhang, J.~Luckas, C.~Chen, R.~Mazzarello, B.~Bichara, M.~Wuttig, Aging mechanism of amorphous phase change materials, Nat. Commun. 6 (2015) 7467.

\bibitem{Gabardi2015}
S.~Gabardi, S.~Caravati, G.~C. Sosso, J.~Behler, M.~Bernasconi, Microscopic origin of resistance drift in the amorphous state of the phase-change compound {G}e{T}e, Phys. Rev. B 92 (2015) 054201.

\bibitem{Zipoli2016}
F.~Zipoli, D.~Krebs, A.~Curioni, {Structural origin of resistance drift in amorphous GeTe}, Phys. Rev. B 93 (2016) 115201.

\bibitem{noe-drift}
P.~Noé, C.~Sabbione, N.~Castellani, G.~Veux, G.~Navarro, V.~Sousa, F.~Hippert, F.~d'Acapito, Structural change with the resistance drift phenomenon in amorphous {GeTe} phase change materials thin films, J. Phys. D: Appl. Phys. 49 (2015) 035305.

\bibitem{caravati2009first}
S.~Caravati, M.~Bernasconi, T.~K{\"u}hne, M.~Krack, M.~Parrinello, First-principles study of crystalline and amorphous {G}e$_2${S}b$_2${T}e$_5$ and the effects of stoichiometric defects, J. Phys. Condens. Matter 21 (2009) 255501.

\bibitem{caravatiblyp}
S.~Caravati, M.~Bernasconi, Influence of the exchange and correlation functional on the structure of amorphous {Ge$_2$Sb$_2$Te$_5$}, Phys. Status Solidi B 252 (2015) 260–266.

\bibitem{Elliott2019}
K.~Konstantinou, F.~C. Mocanu, T.-H. Lee, S.~R. Elliott, {Revealing the intrinsic nature of the mid-gap defects in amorphous Ge$_2$Sb$_2$Te$_5$}, Nat. Commun. 10 (2019) 3065.

\bibitem{Kostantinou2022}
K.~Konstantinou, J.~Akola, S.~R. Elliott, {Inherent electron and hole trapping in amorphous phase-change memory materials: Ge$_2$Sb$_2$Te$_5$}, J. Mater. Chem. 10 (2022) 6744.

\bibitem{Kostantinou2023}
K.~Konstantinou, S.~R. Elliott, Atomistic modeling of charge-trapping defects in amorphous {Ge-Sb-Te} phase-change memory materials, Phys. Status Solidi RRL 17 (2023) 2200496.

\bibitem{mocanu2018modeling}
F.~C. Mocanu, K.~Konstantinou, T.~H. Lee, N.~Bernstein, V.~L. Deringer, G.~Cs{\'a}nyi, S.~R. Elliott, Modeling the phase-change memory material, {G}e$_2${S}b$_2${T}e$_5$, with a machine-learned interatomic potential, J. Phys. Chem. B 122 (2018) 8998--9006.

\bibitem{Massobrio2023}
M.~Guerboub, S.~D.~W. Wendji, C.~Massobrio, A.~Bouzid, M.~Boero, G.~Ori, E.~Martin, Impact of the local atomic structure on the thermal conductivity of amorphous {G}e$_2${S}b$_2${T}e$_5$, J. Chem. Phys. 158 (2023) 084504.

\bibitem{omar2024}
O.~Abou El~Kheir, L.~Bonati, M.~Parrinello, M.~Bernasconi, Unraveling the crystallization kinetics of the {G}e$_2${S}b$_2${T}e$_5$ phase change compound with a machine-learned interatomic potential, npj Comput. Mater. 10 (2024) 33.

\bibitem{acharya2025}
D.~Acharya, O.~Abou El~Kheir, S.~Marcorini, M.~Bernasconi, Simulation of the crystallization kinetics of {G}e$_2${S}b$_2${T}e$_5$ nanoconfined in superlattice geometries for phase change memories, Nanoscale 17 (2025) 13828--13841.

\bibitem{hse06}
A.~V. Krukau, O.~A. Vydrov, A.~F. Izmaylov, G.~E. Scuseria, Influence of the exchange screening parameter on the performance of screened hybrid functionals, J. Chem. Phys. 125 (2006) 224106.

\bibitem{laio}
A.~Laio, M.~Parrinello, Escaping free-energy minima, Proc. Natl. Acad. Sci. U.S.A. 99 (2002) 12562--12566.

\bibitem{laiogervasio}
F.~Gervasio, A.~Laio, Metadynamics: a method to simulate rare events and reconstruct the free energy in biophysics, chemistry and material science, Rep. Prog. Phys. 71 (2002) 126601.

\bibitem{barducci}
A.~Barducci, M.~Bonomi, M.~Parrinello, Metadynamics, Wiley Interdiscip. Rev. Comput. Mol. Sci. 1 (2011) 826--843.

\bibitem{DeePMD4}
L.~Zhang, J.~Han, H.~Wang, R.~Car, W.~E, {Deep Potential Molecular Dynamics: A Scalable Model with the Accuracy of Quantum Mechanics}, Phys. Rev. Lett. 120 (2018) 143001.

\bibitem{DeePMD2}
J.~Zeng, D.~Zhang, D.~Lu, P.~Mo, et~al., {DeePMD-kit v2: A software package for deep potential models}, J. Chem. Phys. 159~(5) (2023) 054801.

\bibitem{DeePMD3}
H.~Wang, L.~Zhang, J.~Han, W.~E, {DeePMD-kit: A deep learning package for many-body potential energy representation and molecular dynamics}, Comput. Phys. Commun. 228 (2018) 178.

\bibitem{PBE}
J.~P. Perdew, K.~Burke, M.~Ernzerhof, Generalized gradient approximation made simple, Phys. Rev. Lett. 77 (1996) 3865.

\bibitem{marcorini}
M.~Marcorini, R.~Pomodoro, O.~Abou El~Kheir, M.~Bernasconi, Viscosity, breakdown of {Stokes-Einstein} relation and dynamical heterogeneity in supercooled liquid {Ge$_2$Sb$_2$Te$_5$} from simulations with a neural network potential, J. Chem. Phys. 163 (2025) 154501.

\bibitem{Omar2025}
O.~Abou El~Kheir, M.~Bernasconi, Million-atom simulation of the set process in phase change memories at the real device scale, Adv. Electr. Mater. 11 (2025) e2500110.

\bibitem{LAMMPS}
A.~P. Thompson, H.~M. Aktulga, R.~Berger, D.~S. Bolintineanu, W.~M. Brown, P.~S. Crozier, P.~J. in~'t Veld, A.~Kohlmeyer, S.~G. Moore, T.~D. Nguyen, R.~Shan, M.~J. Stevens, J.~Tranchida, C.~Trott, S.~J. Plimpton, {LAMMPS} - a flexible simulation tool for particle-based materials modeling at the atomic, meso, and continuum scales, Comp. Phys. Comm. 271 (2022) 108171.

\bibitem{noseart}
S.~Nos\'e, A unified formulation of the constant temperature molecular-dynamics methods, J. Chem. Phys. 8 (1984) 511--519.

\bibitem{hoover}
W.~J. Hoover, Canonical dynamics: Equilibrium phase-space distributions, Phys. Rev. A 31 (1985) 1695--1697.

\bibitem{quickstep1}
T.~Kuehne, et~al., {CP2K}: An electronic structure and molecular dynamics software package - quickstep: Efficient and accurate electronic structure calculations, J. Chem. Phys. 152 (2020) 194103.

\bibitem{quickstep2}
J.~Vandevondele, M.~Krack, F.~Mohamed, M.~Parrinello, T.~Chassaing, J.~Hutter, {Quickstep: Fast and accurate density functional calculations using a mixed Gaussian and plane waves approach}, Comput. Phys. Commun. 167 (2005) 103--128.
\newblock \href {https://doi.org/10.1016/j.cpc.2004.12.014. www.cp2k.org} {\path{doi:10.1016/j.cpc.2004.12.014. www.cp2k.org}}.

\bibitem{GTH2}
M.~Krack, {Pseudopotentials for H to Kr optimized for gradient-corrected exchange-correlation functionals}, Theor. Chem. Acc. 114 (2005) 145--152.

\bibitem{GTH1}
S.~Goedecker, M.~Teter, J.~Hutter, {Separable dual-space Gaussian pseudopotentials}, Phys. Rev. B 54 (1996) 1703.

\bibitem{ghezzi}
G.~E. Ghezzi, J.~Y. Raty, S.~Maitrejean, A.~Roule, E.~Elkaim, F.~Hippert, Effect of carbon doping on the structure of amorphous {GeTe} phase change material, Appl. Phys. Lett. 99 (2011) 151906.

\bibitem{qparam}
J.~R. Errington, P.~G. Debenedetti, Relationship between structural order and the anomalies of liquid water, Nature 409 (2001) 318--321.

\bibitem{spreafico}
E.~Spreafico, S.~Caravati, M.~Bernasconi, First-principles study of lisquid and amorphous {I}n{G}e{T}e$_2$, Phys. Rev. B 84 (2011) 144205.

\bibitem{Abelson}
B.-S. Lee, J.~R. Abelson, S.~G. Bishop, D.-H. Kang, B.-k. Cheong, K.-B. Kim, Investigation of the optical and electronic properties of {Ge$_2$Sb$_2$Te$_5$} phase change material in its amorphous, cubic, and hexagonal phases, J. Appl. Phys. 97 (2005) 093509.

\bibitem{Tian2023}
J.~Tian, W.~Ma, P.~Boulet, M.~Record, Electronic and transport properties of strained and unstrained {Ge$_2$Sb$_2$Te$_5$}: A {DFT} investigation, Materials 16 (2023) 5015.

\bibitem{Kaes2016}
M.~Kaes, M.~Salinga, Impact of defect occupation on conduction in amorphous {Ge$_2$Sb$_2$Te$_5$}, Sci. Rep. 16 (2016) 31699.

\bibitem{InfoMatGeSe}
M.~Xu, M.~Xu, X.~Miao, Deep machine learning unravels the structural origin of mid-gap states in chalcogenide glass for high-density memory integration, InfoMat 4 (2022) e12315.
\newblock \href {https://doi.org/https://doi.org/10.1002/inf2.12315} {\path{doi:https://doi.org/10.1002/inf2.12315}}.

\bibitem{lee2025wrongbonds}
T.~H. Lee, Chemical ordering in liquid and supercooled {Ge$_2$Sb$_2$Te$_5$} phase-change materials, materials 18 (2025) 3900.

\bibitem{bonomi}
M.~Bonomi, M.~Parrinello, Enhanced sampling in the well-tempered ensemble, Phys. Rev. Lett. 104 (2010) 190601.

\bibitem{refTauc}
J.~Tauc, R.~Grigorovici, A.~Vancu, Optical properties and electronic structure of amorphous germanium, Phys. Status Solidi B 15 (1966) 627--637.
\newblock \href {https://doi.org/https://doi.org/10.1002/pssb.19660150224} {\path{doi:https://doi.org/10.1002/pssb.19660150224}}.

\bibitem{slassi}
A.~Slassi, L.-S. Medondjio, A.~Padovani, F.~Tavanti, X.~He, S.~Clima, D.~Garbin, B.~Kaczer, L.~Larcher, P.~Ordejón, A.~Calzolari, Device-to-materials pathway for electron traps detection in amorphous {GeSe-Based Selectors}, Adv. Electron. Mater. 23 (2024) 2201224.

\bibitem{CaravatiSSS}
S.~Caravati, D.~Baratella, P.~Fantini, M.~Bernasconi, In-gap electronic states of {GeAsSe} and {SiGeAsSe} alloys for selector devices from atomistic simulations, Solid State Sciences 170 (2025) 108127.

\end{thebibliography}


\begin{thebibliography}{1}
\expandafter\ifx\csname url\endcsname\relax
  \def\url#1{\texttt{#1}}\fi
\expandafter\ifx\csname urlprefix\endcsname\relax\def\urlprefix{URL }\fi
\expandafter\ifx\csname href\endcsname\relax
  \def\href#1#2{#2} \def\path#1{#1}\fi

\bibitem{hse06}
A.~V. Krukau, O.~A. Vydrov, A.~F. Izmaylov, G.~E. Scuseria, Influence of the exchange screening parameter on the performance of screened hybrid functionals, J. Chem. Phys. 125 (2006) 224106.

\bibitem{PBE}
J.~P. Perdew, K.~Burke, M.~Ernzerhof, Generalized gradient approximation made simple, Phys. Rev. Lett. 77 (1996) 3865.

\bibitem{caravatiblyp}
S.~Caravati, M.~Bernasconi, Influence of the exchange and correlation functional on the structure of amorphous {Ge$_2$Sb$_2$Te$_5$}, Phys. Status Solidi B 252 (2015) 260–266.

\end{thebibliography}

\medskip
\noindent
\textbf{Supplementary Material} \par
The online version contains supplementary material
available at DOI:.....

\medskip
\noindent
\textbf{Data availability} \par {The atomic configurations of all models are available in the Materials Cloud repository at  https://doi.org/......

\medskip
\noindent
\textbf{Code Availability Statement} \par
LAMMPS, DeePMD and CP2k  are free and open source codes available at https:// lammps.sandia.gov,  http://www.deepmd.org and http://www.cp2k.org, respectively.  The NN potential and the DFT database used for its generation, as reported in Ref. \cite{omar2024}, are available in the Materials Cloud repository at https://doi.org/10.24435/materialscloud:a8-45.
\medskip

\noindent
\textbf{Acknowledgments} \par 
The project has received funding from European Union NextGenerationEU through the Italian Ministry of
University and Research under PNRR M4C2I1.4 ICSC 
Centro Nazionale di Ricerca in High Performance Computing, Big Data and Quantum Computing (Grant No.
CN00000013). We acknowledge the CINECA award under the ISCRA initiative for the availability of high-performance computing resources and support.

\medskip
\noindent
\textbf{Competing interests} \par
The authors declare no conflicts of interest.

\medskip
\noindent
\textbf{Authors contributions} \par
The authors equally contributed to the work.

\end{document}

\maketitle

\tableofcontents
\section{Electronic DOS of NN models}

In Figs. S1-S30 we show the electronic density of states (DOS) computed with the HSE hybrid functional \cite{hse06} of the thirty 297-atom amorphous models of GST generated by quenching the melt in NN-MD and then geometry-optimized at zero temperature with the HSE functional. Isosurfaces of the most localized in-gap states analyzed in the article are also shown (see Table S2).

\begin{figure}[H]
    \centering
        \renewcommand\figurename{Figure~S$\!\!$}
    \includegraphics{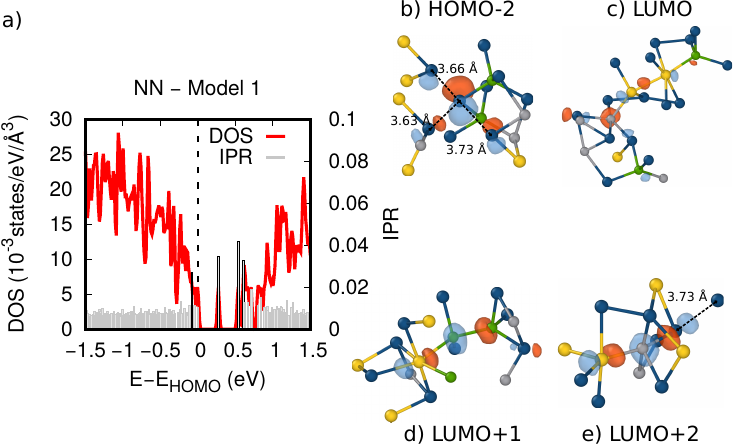}
    \caption{a) A zoom of the electronic DOS of model NN-1 (red curve) close to the band gap and the value of the IPR (gray spikes) of KS states at the $\Gamma$ point. The zero of energy is the highest occupied state (HOMO). The isosurfaces of states with high IPR (highlighted with black border) are shown in panels b)-e). b) The HOMO-2 is localized on a 2-coordinated Te atom bonded to two Ge$_{\rm T}$ atoms and forming there long Te-Te contacts (3.63- 3.73 \AA). c) The LUMO is localized close to a 5-coordinated Ge atom with one wrong bond with a partial contribution of a small chain of four wrong bonds (Ge-Sb-Sb-Ge$_{\rm T}$). d) The LUMO+1 is localized in the neighborhood of two bonded Ge$_{\rm T}$ atoms and an overcoordinated Sb atom. e) LUMO+2 is localized near a 5-coordinated Ge atom with two wrong bonds. In all panels, Ge, Ge$_{\rm T}$ (tetrahedral Ge), Sb and Te are depicted with gray, green, yellow and blue spheres.}
    \label{fig:NN1}
\end{figure}

\begin{figure}[H]
    \centering
           \renewcommand\figurename{Figure~S$\!\!$}
    \includegraphics{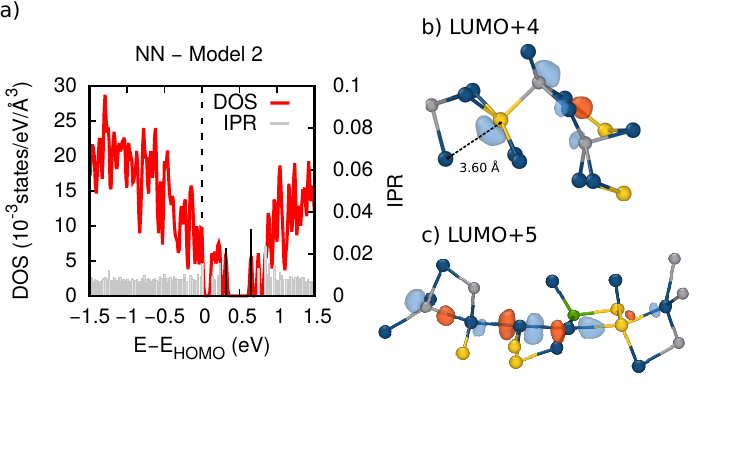}
    \caption{a) A zoom of the electronic DOS of model NN-2 (red curve) close to the band gap and the value of the IPR (gray spikes) of KS states at the $\Gamma$ point. The zero of energy is the highest occupied state (HOMO). The isosurfaces of states with high IPR (highlighted with black border) are shown in panels b and c. b) The LUMO+4 is localized in the neighborhood  of a wrong bond between a 5-coordinated Sb and a 4-coordinated Ge atom. The Sb atom also forms a long contact with a Te atom (3.60 \AA). c) The LUMO+5 is localized on a chain of Te atoms. In all panels, Ge, Ge$_{\rm T}$ (tetrahedral Ge), Sb and Te are depicted with gray, green, yellow and blue spheres.}
    \label{fig:NN2}
\end{figure}

\begin{figure}[H]
    \centering
           \renewcommand\figurename{Figure~S$\!\!$}
    \includegraphics{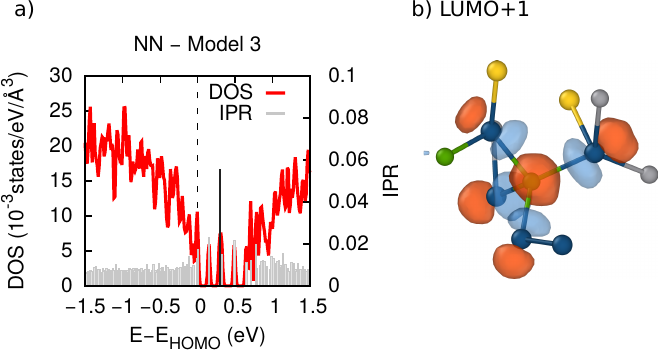}
    \caption{a) A zoom of the electronic DOS of model NN-3 (red curve) close to the band gap and the value of the IPR (gray spikes) of KS states at the $\Gamma$ point. The zero of energy is the highest occupied state (HOMO).   The isosurfaces of the state with highest IPR (highlighted with black border) is shown in panel b. b) The LUMO+1 is localized on a Ge$_{\rm T}$ bonded to four Te atoms. Ge, Ge$_{\rm T}$ (tetrahedral Ge), Sb and Te are depicted with gray, green, yellow and blue spheres.}
    \label{fig:NN3}
\end{figure}

\begin{figure}[H]
    \centering
           \renewcommand\figurename{Figure~S$\!\!$}
    \includegraphics{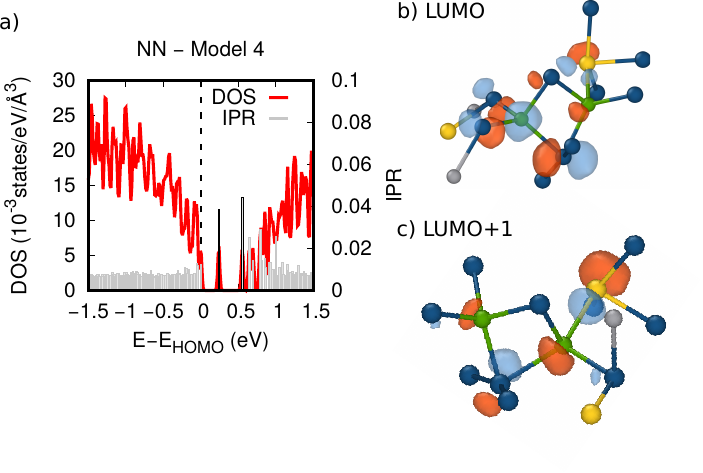}
    \caption{a) A zoom of the electronic DOS of model NN-4 (red curve) close to the band gap and the value of the IPR (gray spikes) of KS states at the $\Gamma$ point. The zero of energy is the highest occupied state (HOMO). The isosurfaces of states with high IPR (highlighted with black border) are shown in panels b and c. b) The LUMO is localized close to a couple of Ge$_{\rm T}$ atoms bonded mainly to Te atoms and one wrong Ge-Sb bond. c) The LUMO+1 is localized on the same environment of the LUMO state.   Ge, Ge$_{\rm T}$ (tetrahedral Ge), Sb and Te are depicted with gray, green, yellow and blue spheres.}
    \label{fig:NN4}
\end{figure}

\begin{figure}[H]
    \centering
           \renewcommand\figurename{Figure~S$\!\!$}
    \includegraphics{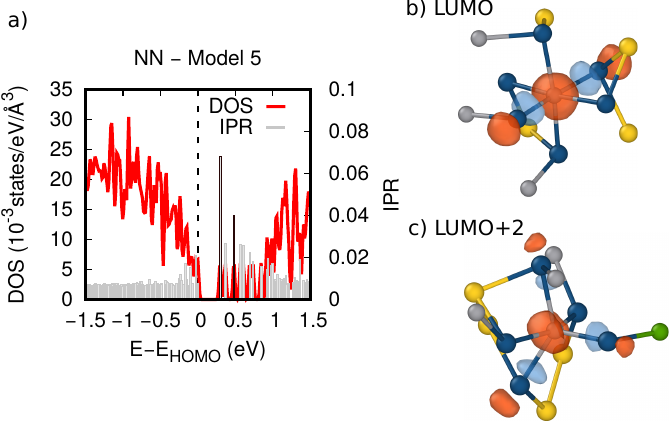}
    \caption{a) A zoom of the electronic DOS of model NN-5 (red curve) close to the band gap and the value of the IPR (gray spikes) of KS states at the $\Gamma$ point. The zero of energy is the highest occupied state (HOMO). The isosurfaces of states with high IPR (highlighted with black border) are shown in panels b and c. The LUMO (panel b) and the LUMO+2 (panel c) are localized on overcoordinated Ge atoms. Ge, Ge$_{\rm T}$ (tetrahedral Ge), Sb and Te are depicted with gray, green, yellow and blue spheres.}
    \label{fig:NN5}
\end{figure}

\begin{figure}[H]
    \centering
           \renewcommand\figurename{Figure~S$\!\!$}
    \includegraphics{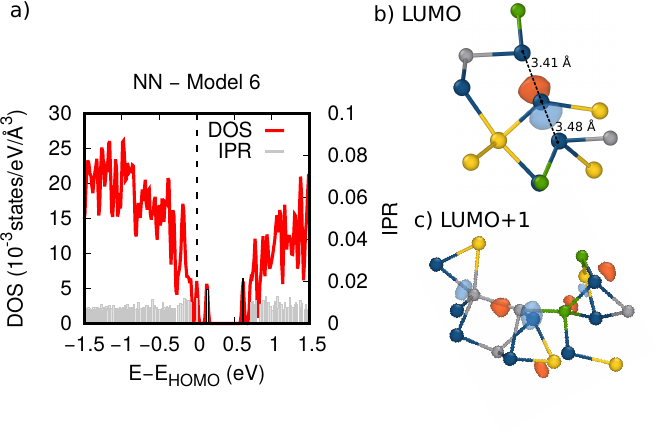}
    \caption{a) A zoom of the electronic DOS of model NN-6 (red curve) close to the band gap and the value of the IPR (gray spikes) of KS states at the $\Gamma$ point. The zero of energy is the highest occupied state (HOMO). The isosurfaces of states with high IPR (highlighted with black border) are shown in panels b and c. b) The LUMO is weakly localized on a 2-coordinated Te atoms forming two long  bonds with Te atoms. c) The LUMO+1 is localized on a long chain of Ge atoms involving overcoordinated  and tetrahedral atoms. Ge, Ge$_{\rm T}$ (tetrahedral Ge), Sb and Te are depicted with gray, green, yellow and blue spheres.}
    \label{fig:NN6}
\end{figure}

\begin{figure}[H]
    \centering
           \renewcommand\figurename{Figure~S$\!\!$}
    \includegraphics{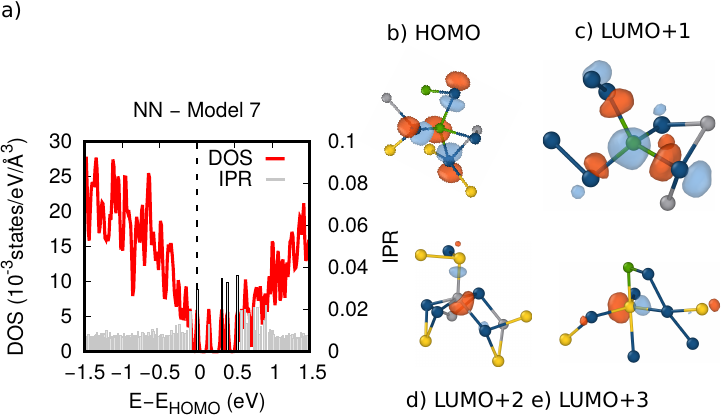}
    \caption{a) A zoom of the electronic DOS of model NN-7 (red curve) close to the band gap and the value of the IPR (gray spikes) of KS states at the $\Gamma$ point. The zero of energy is the highest occupied state (HOMO). The isosurfaces of states with high IPR (highlighted with black border) are shown in panels b)-e). b) The HOMO state which is localized on a 4-coordinated Ge atom with a q-order parameter of 0.78 very close to that of a G$_{\rm T}$ atom. c) The LUMO+1 is localized on a Ge$_{\rm T}$ atom. d) The LUMO+2 is localized on a 5-coordinated Ge atom with a wrong bond. e) LUMO+3 is localized on a 5-coordinated Sb atom bonded to a Ge$_{\rm T}$ atom. Ge, Ge$_{\rm T}$ (tetrahedral Ge), Sb and Te are depicted with gray, green, yellow and blue spheres.}
    \label{fig:NN7}
\end{figure}

\begin{figure}[H]
    \centering
           \renewcommand\figurename{Figure~S$\!\!$}
    \includegraphics{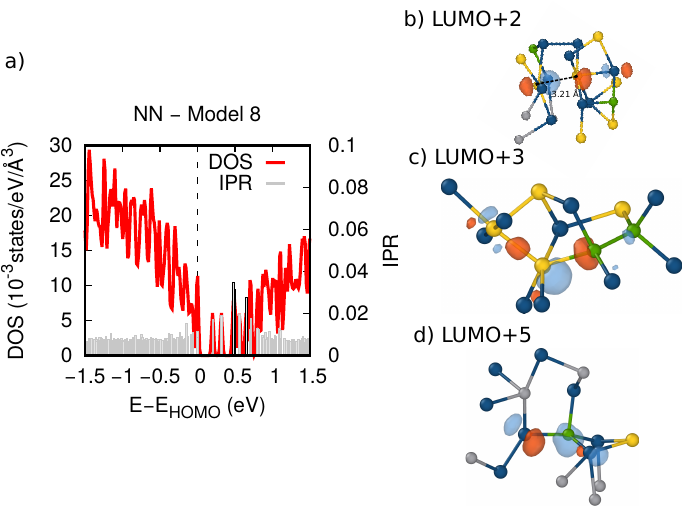}
    \caption{a) A zoom of the electronic DOS of model NN-8 (red curve) close to the band gap and the value of the IPR (gray spikes) of KS states at the $\Gamma$ point. The zero of energy is the highest occupied state (HOMO). The isosurfaces of states with high IPR (highlighted with black border) are shown in panels b)-d). b) The LUMO+2 is localized on two Sb atoms forming a bond slightly longer than the cutoff. c) The LUMO+3 is localized on a chain of wrong  Sb-Sb and Sb-Ge$_{\rm T}$ wrong bonds. d) The LUMO+5 is  on a Ge$_{\rm T}$ atom. Ge, Ge$_{\rm T}$ (tetrahedral Ge), Sb and Te are depicted with gray, green, yellow and blue spheres.}
    \label{fig:NN8}
\end{figure}

\begin{figure}[H]
    \centering
           \renewcommand\figurename{Figure~S$\!\!$}
    \includegraphics{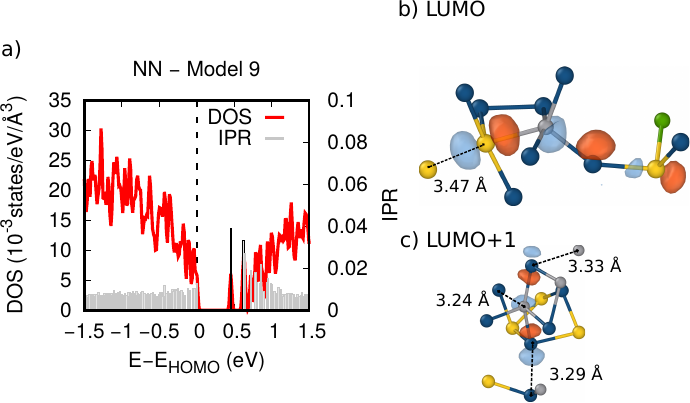}
    \caption{a) A zoom of the electronic DOS of model NN-9 (red curve) close to the band gap and the value of the IPR (gray spikes) of KS states at the $\Gamma$ point. The zero of energy is the highest occupied state (HOMO). The isosurfaces of states with high IPR (highlighted with black border) are shown in panels b, and c. b) The LUMO is localized close to an overcoordinated Ge atom with a Sb-Ge wrong bond and on an Sb atom forming a bond slightly longer than the cutoff. c) The LUMO+1 is localized on an a Ge-atom in a defective octahedral geometry. Ge, Ge$_{\rm T}$ (tetrahedral Ge), Sb and Te are depicted with gray, green, yellow and blue spheres.}
    \label{fig:NN9}
\end{figure}

\begin{figure}[H]
    \centering
           \renewcommand\figurename{Figure~S$\!\!$}
    \includegraphics{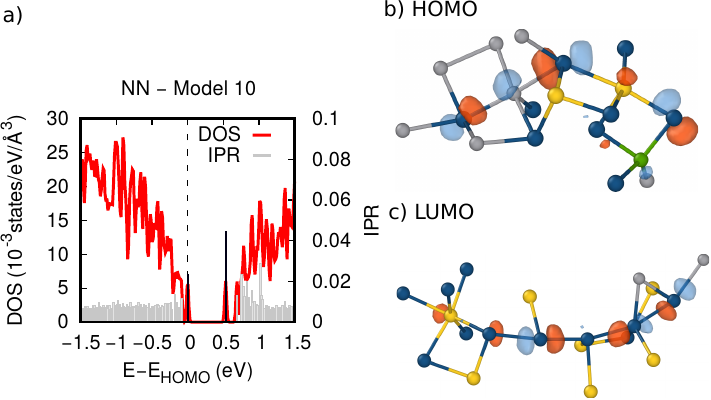}
    \caption{a) A zoom of the electronic DOS of model NN-10 (red curve) close to the band gap and the value of the IPR (gray spikes) of KS states at the $\Gamma$ point. The zero of energy is the highest occupied state (HOMO). The isosurfaces of states with high IPR (highlighted with black border) are shown in panels b, and c. b) The HOMO is localized on  the neighborhood of an overcoordinated Ge  and Sb atoms in a defective octahedral geometry. c) The LUMO is localized on a long chain of Te-Te bonds. Ge, Ge$_{\rm T}$ (tetrahedral Ge), Sb and Te are depicted with gray, green, yellow and blue spheres.}
    \label{fig:NN10}
\end{figure}

\begin{figure}[H]
    \centering
           \renewcommand\figurename{Figure~S$\!\!$}
    \includegraphics{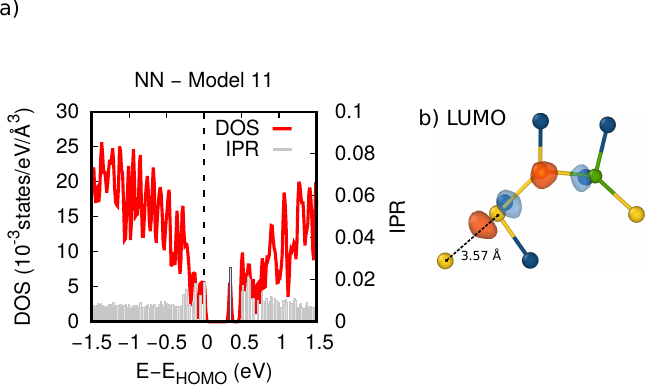}
    \caption{a) A zoom of the electronic DOS of model NN-11 (red curve) close to the band gap and the value of the IPR (gray spikes) of KS states at the $\Gamma$ point. The zero of energy is the highest occupied state (HOMO).   The isosurfaces of the state with highest IPR (highlighted with black border) is shown in panel b. b) The LUMO is localized on a chain of Sb-Sb-Ge$_{\rm T}$-Sb wrong bonds. Ge, Ge$_{\rm T}$ (tetrahdral Ge), Sb and Te are depicted with gray, green, yellow and blue spheres.}
    \label{fig:NN11}
\end{figure}

\begin{figure}[H]
    \centering
           \renewcommand\figurename{Figure~S$\!\!$}
    \includegraphics{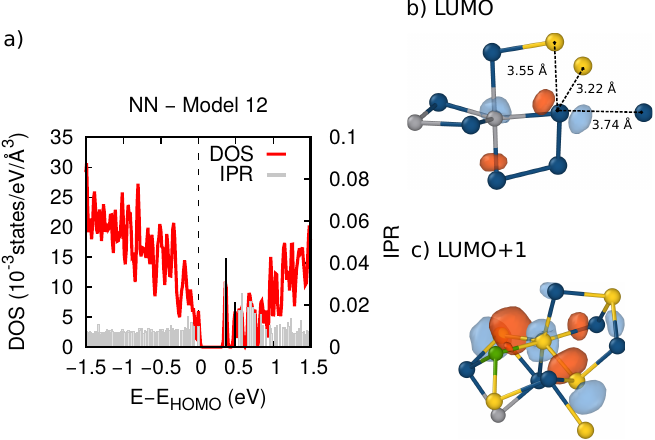}
    \caption{a) A zoom of the electronic DOS of model NN-12 (red curve) close to the band gap and the value of the IPR (gray spikes) of KS states at the $\Gamma$ point. The zero of energy is the highest occupied state (HOMO). The isosurfaces of states with high IPR (highlighted with black border) are shown in panels b, and c. b) The LUMO is localized on a 5-coordinated Ge atom bonded to a 2-coordinated Te atom. The latter forms three long contacts with Sb and Te atoms. c) The LUMO+1 is localized on an overcoordinated Sb atom with several wrong bonds. Ge, Ge$_{\rm T}$ (tetrahdral Ge), Sb and Te are depicted with gray, green, yellow and blue spheres.}
    \label{fig:NN12}
\end{figure}

\begin{figure}[H]
    \centering
           \renewcommand\figurename{Figure~S$\!\!$}
    \includegraphics{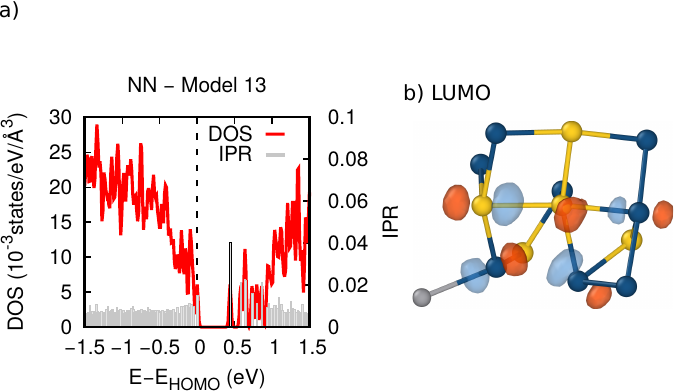}
    \caption{a) A zoom of the electronic DOS of model NN-13 (red curve) close to the band gap and the value of the IPR (gray spikes) of KS states at the $\Gamma$ point. The zero of energy is the highest occupied state (HOMO).   The isosurfaces of the state with highest IPR (highlighted with black border) is shown in panel b. b) The LUMO is localized close to a 5-coordinated Sb atom bonded to two Sb atoms. Ge, Ge$_{\rm T}$ (tetrahdral Ge), Sb and Te are depicted with gray, green, yellow and blue spheres.}
    \label{fig:NN13}
\end{figure}

\begin{figure}[H]
    \centering
           \renewcommand\figurename{Figure~S$\!\!$}
    \includegraphics{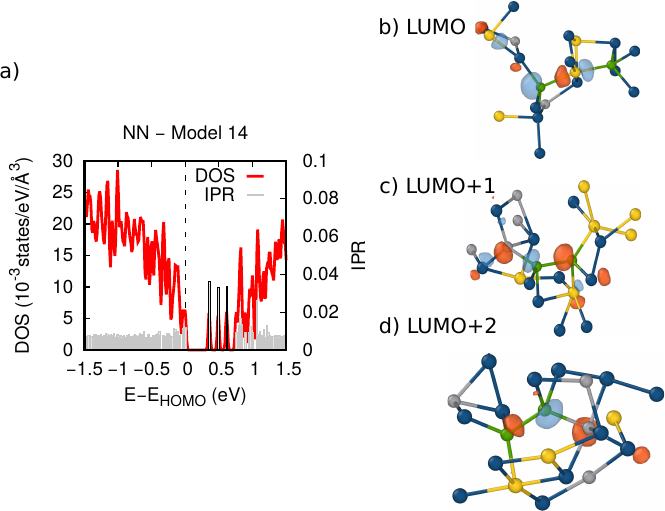}
    \caption{a) A zoom of the electronic DOS of model NN-14 (red curve) close to the band gap and the value of the IPR (gray spikes) of KS states at the $\Gamma$ point. The zero of energy is the highest occupied state (HOMO). The isosurfaces of states with high IPR (highlighted with black border) are shown in panels b, c, and d. b) The LUMO is localized on a 5-coordinated Sb bonded to two Ge$_{\rm T}$ atoms. The LUMO+1 (panel c) and the LUMO+2 (panel d) are localized on a chain of Ge atoms with two Ge$_{\rm T}$ atoms. Ge, Ge$_{\rm T}$ (tetrahdral Ge), Sb and Te are depicted with gray, green, yellow and blue spheres.}
    \label{fig:NN14}
\end{figure}

\begin{figure}[H]
    \centering
           \renewcommand\figurename{Figure~S$\!\!$}
    \includegraphics{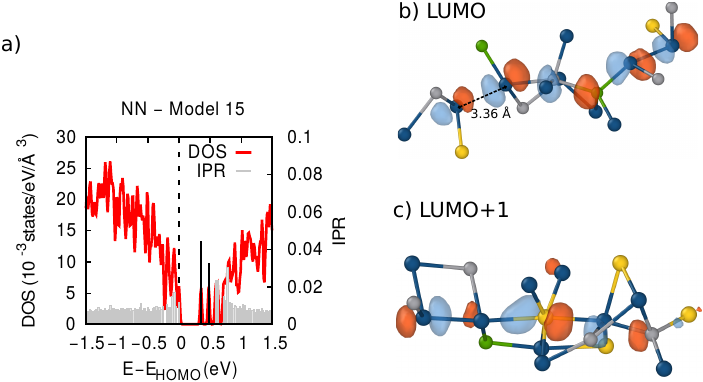}
    \caption{a) A zoom of the electronic DOS of model NN-15 (red curve) close to the band gap and the value of the IPR (gray spikes) of KS states at the $\Gamma$ point. The zero of energy is the highest occupied state (HOMO). The isosurfaces of states with high IPR (highlighted with black border) are shown in panels b, and c. b) The LUMO is localized on long  chain of wrong bonds.  c)  LUMO+1 is localized close to a 6-coordinated Sb atom, a Te-Te wrong bond and a 4-coordinated Te atom forming a Te-Te bond.  Ge, Ge$_{\rm T}$ (tetrahdral Ge), Sb and Te are depicted with gray, green, yellow and blue spheres.}
    \label{fig:NN15}
\end{figure}

\begin{figure}[H]
    \centering
           \renewcommand\figurename{Figure~S$\!\!$}
    \includegraphics{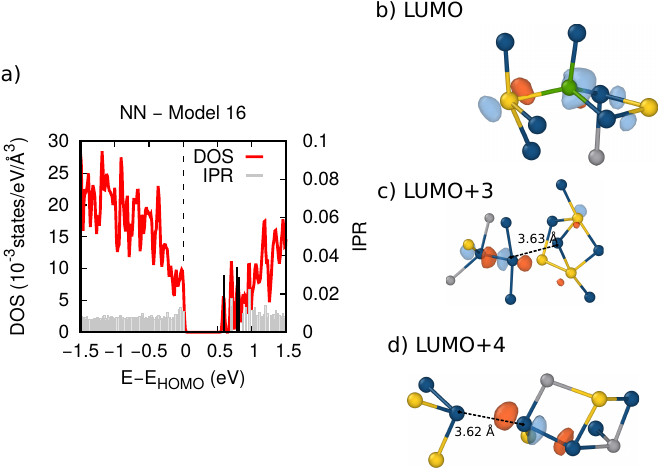}
    \caption{a) A zoom of the electronic DOS of model NN-16 (red curve) close to the band gap and the value of the IPR (gray spikes) of KS states at the $\Gamma$ point. The zero of energy is the highest occupied state (HOMO). The isosurfaces of states with high IPR (highlighted with black border) are shown in panels b, c and d. b) The LUMO is localized on a Ge$_{\rm T}$ atom. LUMO+3 (panel c) and LUMO+4 (panel d) are localized on a 3-coordinated Te atom forming Te-Te wrong bond and a long Te-Te contact.   Ge, Ge$_{\rm T}$ (tetrahdral Ge), Sb and Te are depicted with gray, green, yellow and blue spheres.}
    \label{fig:NN16}
\end{figure}

\begin{figure}[H]
    \centering
           \renewcommand\figurename{Figure~S$\!\!$}
    \includegraphics{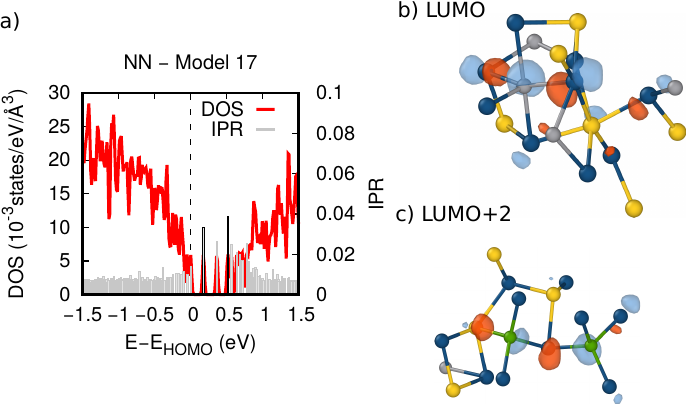}
    \caption{a) A zoom of the electronic DOS of model NN-17 (red curve) close to the band gap and the value of the IPR (gray spikes) of KS states at the $\Gamma$ point. The zero of energy is the highest occupied state (HOMO). The isosurfaces of states with high IPR (highlighted with black border) are shown in panels b and c. b) The LUMO is localized on 5-coordinated Ge atom. c) LUMO+2 is localized on a  chain of two Ge$_{\rm T}$  and a Ge-Sb wrong bond.   Ge, Ge$_{\rm T}$ (tetrahdral Ge), Sb and Te are depicted with gray, green, yellow and blue spheres.}
    \label{fig:NN17}
\end{figure}

\begin{figure}[H]
    \centering
           \renewcommand\figurename{Figure~S$\!\!$}
    \includegraphics{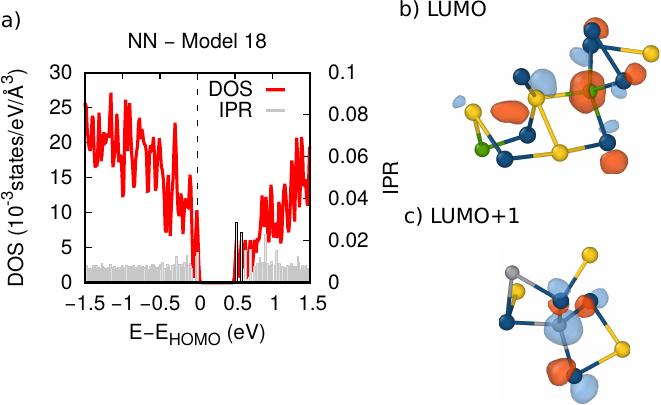}
    \caption{a) A zoom of the electronic DOS of model NN-18 (red curve) close to the band gap and the value of the IPR (gray spikes) of KS states at the $\Gamma$ point. The zero of energy is the highest occupied state (HOMO). The isosurfaces of states with high IPR (highlighted with black border) are shown in panels b and c. b) The LUMO is localized on a Ge$_{\rm T}$ atom with a wrong bond. c) LUMO+1 is localized on a 4-coordinated Ge atom.   Ge, Ge$_{\rm T}$ (tetrahdral Ge), Sb and Te are depicted with gray, green, yellow and blue spheres.}
    \label{fig:NN18}
\end{figure}

\begin{figure}[H]
    \centering
           \renewcommand\figurename{Figure~S$\!\!$}
    \includegraphics{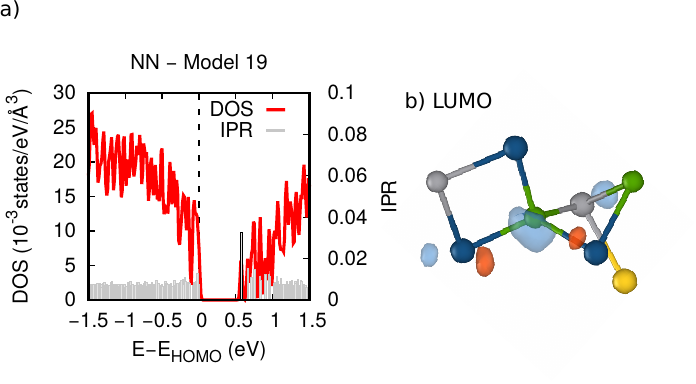}
    \caption{a) A zoom of the electronic DOS of model NN-19 (red curve) close to the band gap and the value of the IPR (gray spikes) of KS states at the $\Gamma$ point. The zero of energy is the highest occupied state (HOMO).   The isosurfaces of the state with highest IPR (highlighted with black border) is shown in panel b. b) The LUMO is localized close to a Ge$_{\rm T}$ atom with a wrong homopolar bond. Ge, Ge$_{\rm T}$, Sb and Te are depicted with gray, green, yellow and blue spheres.}
    \label{fig:NN19}
\end{figure}

\begin{figure}[H]
    \centering
           \renewcommand\figurename{Figure~S$\!\!$}
    \includegraphics{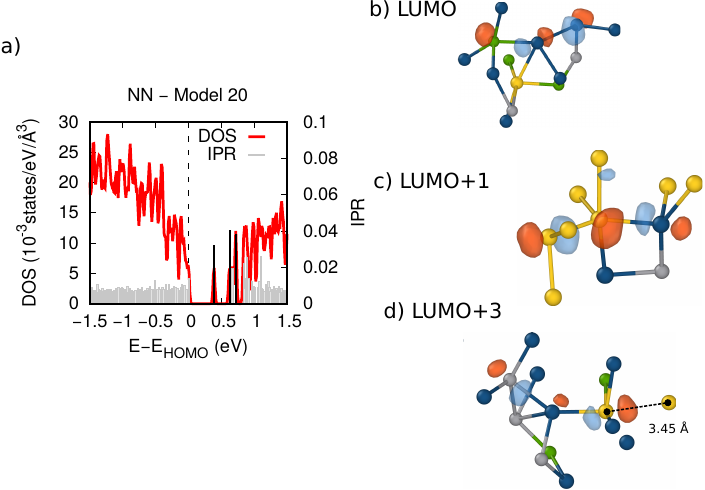}
    \caption{a) A zoom of the electronic DOS of model NN-20 (red curve) close to the band gap and the value of the IPR (gray spikes) of KS states at the $\Gamma$ point. The zero of energy is the highest occupied state (HOMO). The isosurfaces of states with high IPR (highlighted with black border) are shown in panels b, c and d. b) The LUMO is localized on a Te atom forming a homopolar bond and a  bond with a Ge$_{\rm T}$ atom. c) LUMO+1 is localized near a 5-coordinated Sb atom forming homopolar bonds. d) LUMO+3 is localized close to a 4-coordinated  Sb forming a long contact with an Sb atom. Ge, Ge$_{\rm T}$ (tetrahdral Ge), Sb and Te are depicted with gray, green, yellow and blue spheres.}
    \label{fig:NN20}
\end{figure}

\begin{figure}[H]
    \centering
           \renewcommand\figurename{Figure~S$\!\!$}
    \includegraphics{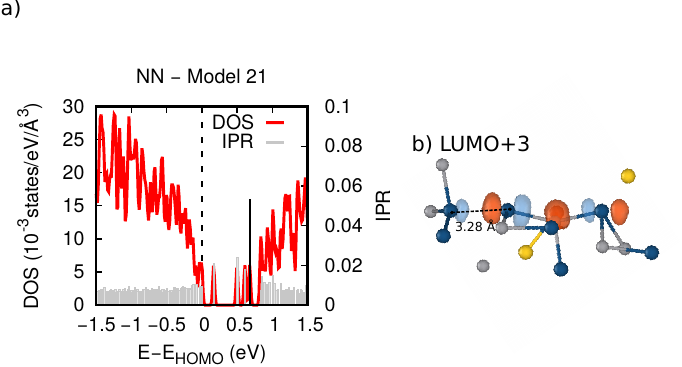}
    \caption{a) A zoom of the electronic DOS of model NN-21 (red curve) close to the band gap and the value of the IPR (gray spikes) of KS states at the $\Gamma$ point. The zero of energy is the highest occupied state (HOMO).   The isosurfaces of the state with highest IPR (highlighted with black border) is shown in panel b. b) The LUMO+3 is localized close to a 5-coordinated  Ge atom bonded to a Te atom forming in turn a long homopolar contact. Ge, Ge$_{\rm T}$ (tetrahdral Ge), Sb and Te are depicted with gray, green, yellow and blue spheres.}
    \label{fig:NN21}
\end{figure}

\begin{figure}[H]
    \centering
           \renewcommand\figurename{Figure~S$\!\!$}
    \includegraphics{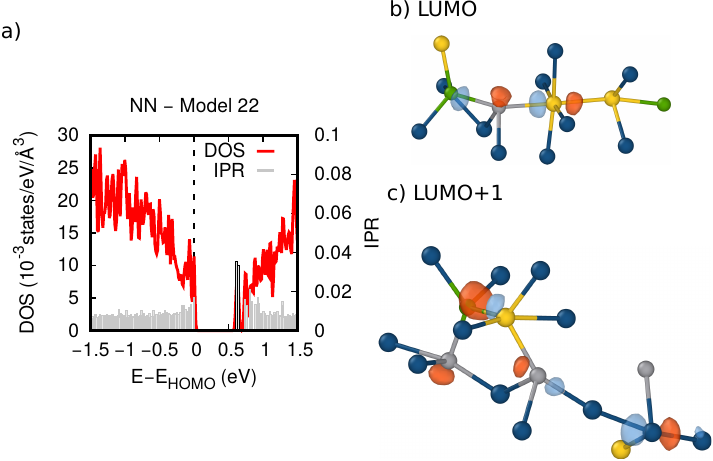}
    \caption{a) A zoom of the electronic DOS of model NN-22 (red curve) close to the band gap and the value of the IPR (gray spikes) of KS states at the $\Gamma$ point. The zero of energy is the highest occupied state (HOMO). The isosurfaces of states with high IPR (highlighted with black border) are shown in panels b and c. LUMO (panel b) and LUMO+1 (panel c) are both localized on a long  chain of wrong bonds. Ge$_{\rm T}$ (tetrahdral Ge), Sb and Te are depicted with gray, green, yellow and blue spheres.}
    \label{fig:NN22}
\end{figure}

\begin{figure}[H]
    \centering
           \renewcommand\figurename{Figure~S$\!\!$}
    \includegraphics{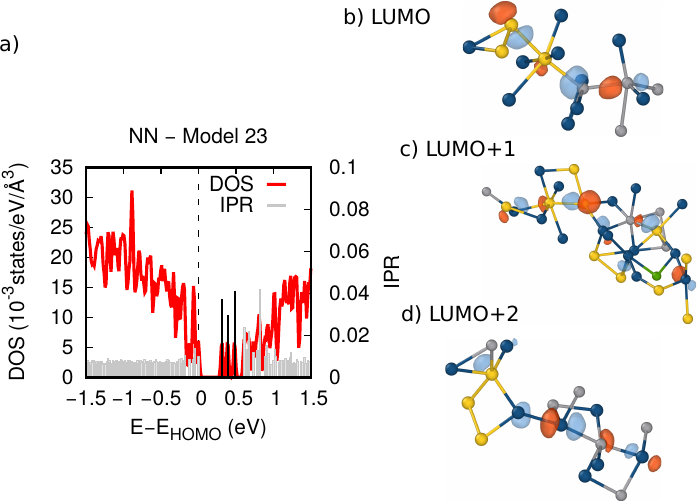}
    \caption{a) A zoom of the electronic DOS of model NN-23 (red curve) close to the band gap and the value of the IPR (gray spikes) of KS states at the $\Gamma$ point. The zero of energy is the highest occupied state (HOMO). The isosurfaces of states with high IPR (highlighted with black border) are shown in panels b, c and d. LUMO (panel b) and LUMO+1 (panel c) are both localized on a long  chain of wrong bonds. d) LUMO+2 is localized close to a 5-coordinated Ge atom bonded to a Te atom forming in turn a homopolar bond. Ge, Ge$_{\rm T}$ (tetrahdral Ge), Sb and Te are depicted with gray, green, yellow and blue spheres.}
    \label{fig:NN23}
\end{figure}

\begin{figure}[H]
    \centering
           \renewcommand\figurename{Figure~S$\!\!$}
    \includegraphics[width=0.8\textwidth]{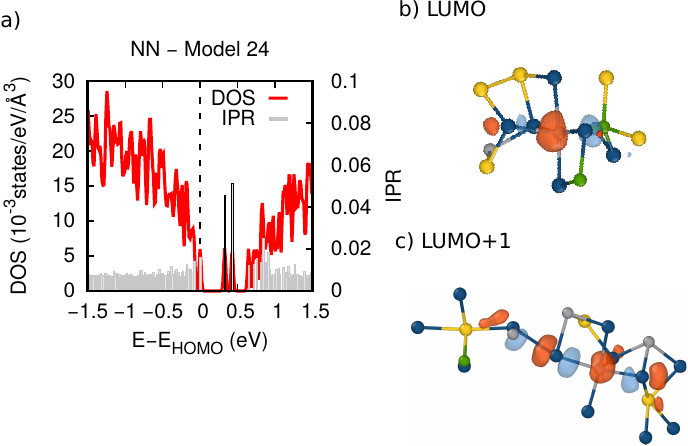}
    \caption{a) A zoom of the electronic DOS of model NN-24 (red curve) close to the band gap and the value of the IPR (gray spikes) of KS states at the $\Gamma$ point. The zero of energy is the highest occupied state (HOMO). The isosurfaces of states with high IPR (highlighted with black border) are shown in panels b and c. b) The LUMO  is localized close to a 6-coordinated Ge atom forming Ge-Ge$_{\rm T}$ bond. c) LUMO+1 is localized close to a 5-coordinated Ge atom bonded to a Te atom forming in turn a homopolar bond. Ge, Ge$_{\rm T}$ (tetrahdral Ge), Sb and Te are depicted with gray, green, yellow and blue spheres.}
    \label{fig:NN24}
\end{figure}

\begin{figure}[H]
    \centering
           \renewcommand\figurename{Figure~S$\!\!$}
    \includegraphics{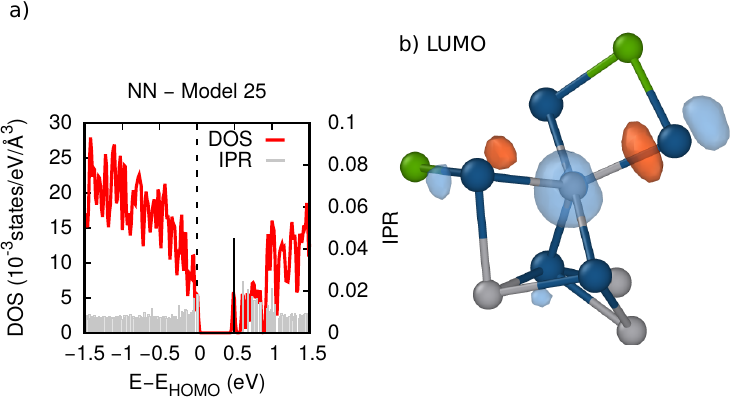}
    \caption{a) A zoom of the electronic DOS of model NN-26 (red curve) close to the band gap and the value of the IPR (gray spikes) of KS states at the $\Gamma$ point. The zero of energy is the highest occupied state (HOMO).   The isosurfaces of the state with highest IPR (highlighted with black border) is shown in panel b. b) The LUMO is localized close to a 5-coordinated Ge atom. Ge, Ge$_{\rm T}$ (tetrahdral Ge), Sb and Te are depicted with gray, green, yellow and blue spheres.}
    \label{fig:NN25}
\end{figure}

\begin{figure}[H]
    \centering
           \renewcommand\figurename{Figure~S$\!\!$}
    \includegraphics{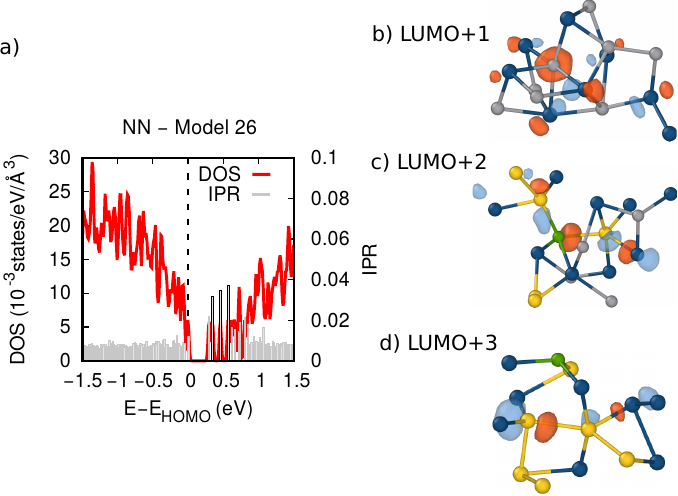}
    \caption{a) A zoom of the electronic DOS of model NN-26 (red curve) close to the band gap and the value of the IPR (gray spikes) of KS states at the $\Gamma$ point. The zero of energy is the highest occupied state (HOMO). The isosurfaces of states with high IPR (highlighted with black border) are shown in panels b, c and d. b) The LUMO+1 state is localized close to a 5-coordinated Ge atom. c) LUMO+2 is localized close to a Ge$_{\rm T}$ atom bonded to two Sb atoms. d) LUMO+3 is localized close to a  5-coordinated Sb atom forming homopolar bonds. Ge, Ge$_{\rm T}$ (tetrahdral Ge), Sb and Te are depicted with gray, green, yellow and blue spheres.}
    \label{fig:NN26}
\end{figure}

\begin{figure}[H]
    \centering
           \renewcommand\figurename{Figure~S$\!\!$}
    \includegraphics{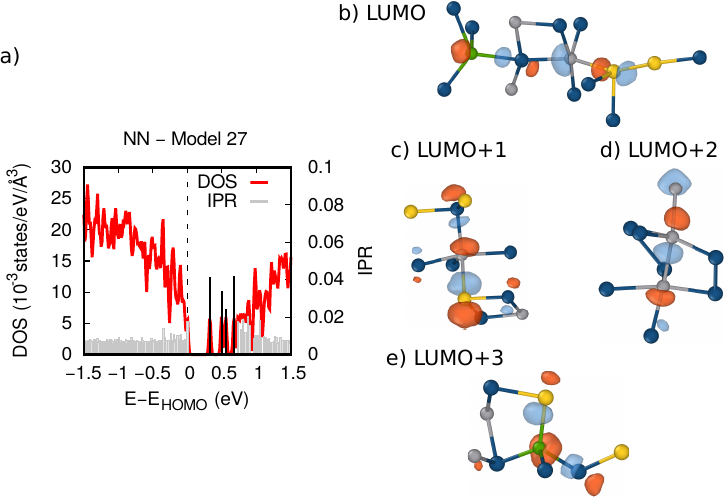}
    \caption{a) A zoom of the electronic DOS of model NN-27 (red curve) close to the band gap and the value of the IPR (gray spikes) of KS states at the $\Gamma$ point. The zero of energy is the highest occupied state (HOMO). The isosurfaces of states with high IPR (highlighted with black border) are shown in panels b)-e). b) LUMO  is localized along a  chain of wrong bonds involving Ge$_{\rm T}$ atom and an overcoordinated Ge atom. c) LUMO+1 is localized on a  a 5-coordinated Ge atom with a wrong bond. d) LUMO+2 is localized on a short chain of Ge atoms. e) LUMO+3 is localized close to  two Ge$_{\rm T}$ atoms forming a wrong bond with an Sb atom. Ge, Ge$_{\rm T}$ (tetrahdral Ge), Sb and Te are depicted with gray, green, yellow and blue spheres.}
    \label{fig:NN27}
\end{figure}

\begin{figure}[H]
    \centering
           \renewcommand\figurename{Figure~S$\!\!$}
    \includegraphics{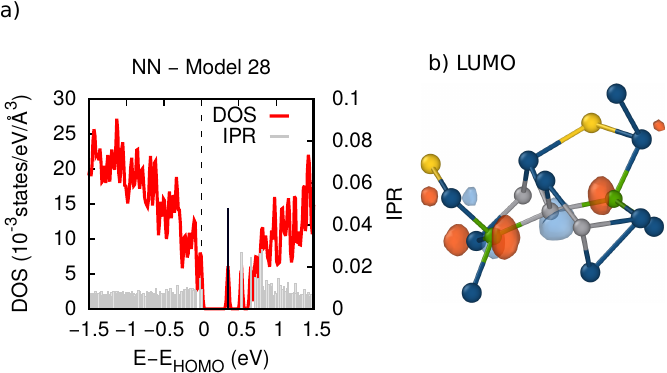}
    \caption{a) A zoom of the electronic DOS of model NN-28 (red curve) close to the band gap and the value of the IPR (gray spikes) of KS states at the $\Gamma$ point. The zero of energy is the highest occupied state (HOMO).   The isosurfaces of the state with highest IPR (highlighted with black border) is shown in panel b. b) The LUMO state is localized on a chain of Ge atoms involving two Ge$_{\rm T}$ atoms. Ge, Ge$_{\rm T}$ (tetrahdral Ge), Sb and Te are depicted with gray, green, yellow and blue spheres.}
    \label{fig:NN28}
\end{figure}

\begin{figure}[H]
    \centering
           \renewcommand\figurename{Figure~S$\!\!$}
    \includegraphics{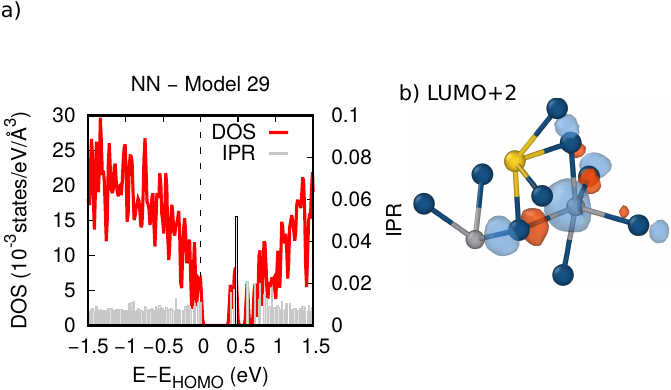}
    \caption{a) A zoom of the electronic DOS of model NN-29 (red curve) close to the band gap and the value of the IPR (gray spikes) of KS states at the $\Gamma$ point. The zero of energy is the highest occupied state (HOMO).   The isosurfaces of the state with highest IPR (highlighted with black border) is shown in panel b. b) The LUMO state is localized close to a 5-coordinated Ge atom. Ge, Ge$_{\rm T}$ (tetrahdral Ge), Sb and Te are depicted with gray, green, yellow and blue spheres.}
    \label{fig:NN29}
\end{figure}

\begin{figure}[H]
    \centering
           \renewcommand\figurename{Figure~S$\!\!$}
    \includegraphics{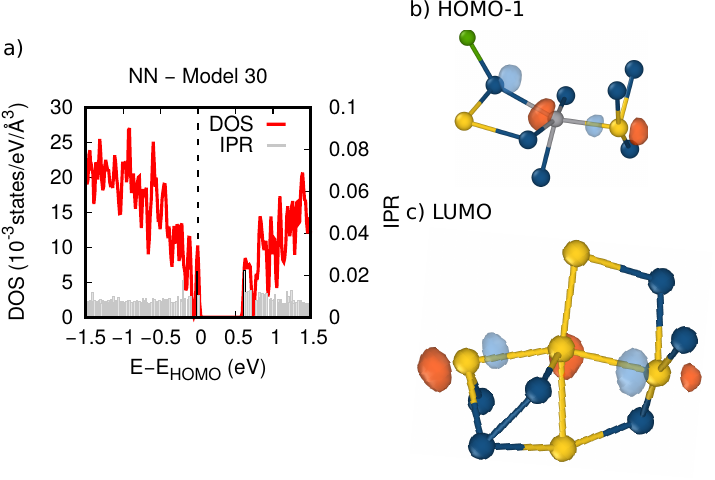}
    \caption{a) A zoom of the electronic DOS of model NN-30 (red curve) close to the band gap and the value of the IPR (gray spikes) of KS states at the $\Gamma$ point. The zero of energy is the highest occupied state (HOMO). The isosurfaces of states with high IPR (highlighted with black border) are shown in panels b and c. b) The HOMO-1 state is localized close to a 5-coordinated Ge atom  forming a wrong Ge-Sb bond. c) LUMO+2 is localized on a chain of Sb atoms. Ge, Ge$_{\rm T}$ (tetrahdral Ge), Sb and Te are depicted with gray, green, yellow and blue spheres.}
    \label{fig:NN30}
\end{figure}

\section{Electronic DOS of DFT models}
In Figs. S31-S40 we show the electronic density of states (DOS) computed with the HSE hydrid functional \cite{hse06} of the ten 216-atom amorphous models of GST generated by quenching the melt in DFT-MD with the PBE functional \cite{PBE} and then geometry-optimized at zero temperature with the HSE functional. Isosurfaces of the most localized in-gap states analyzed in the article are also shown (see Table S2).

\begin{figure}[H]
    \centering
           \renewcommand\figurename{Figure~S$\!\!$}
    \includegraphics{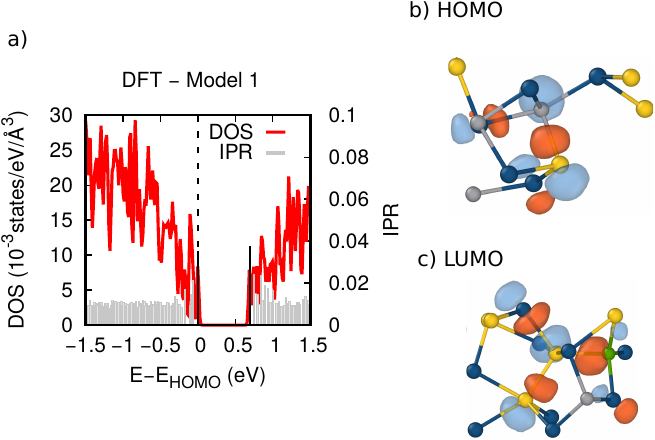}
    \caption{a) A zoom of the electronic DOS of model DFT-1 (red curve) close to the band gap and the value of the IPR (gray spikes) of KS states at the $\Gamma$ point. The zero of energy is the highest occupied state (HOMO). The isosurfaces of states with high IPR (highlighted with black border) are shown in panels b and c. b) The HOMO state is localized close to a 4-coordinated Ge atom forming a wrong bond with a 3-coordinated Sb atom. c) LUMO is localized on a 5-coordinated Sb atom forming wrong bonds with Sb and with  Ge$_{\rm T}$ atoms. Ge, Ge$_{\rm T}$ (tetrahdral Ge), Sb and Te are depicted with gray, green, yellow and blue spheres.}
    \label{fig:DFT1}
\end{figure}

\begin{figure}[H]
    \centering
           \renewcommand\figurename{Figure~S$\!\!$}
    \includegraphics{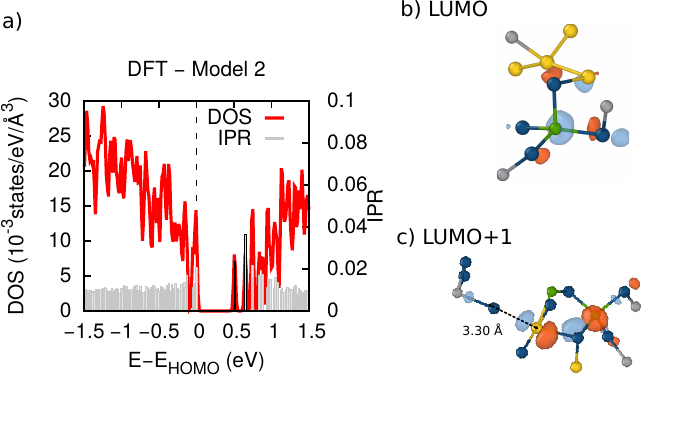}
    \caption{a) A zoom of the electronic DOS of model DFT-2 (red curve) close to the band gap and the value of the IPR (gray spikes) of KS states at the $\Gamma$ point. The zero of energy is the highest occupied state (HOMO). The isosurfaces of states with high IPR (highlighted with black border) are shown in panels b and c. b) The LUMO is localized on a Ge$_{\rm T}$ atom.  c) LUMO+1 is localized close to a Ge$_{\rm T}$ atom and a 4-coordinated Sb atom forming wrong bond with Ge$_{\rm T}$ atoms and a long contact with Te atom. Ge, Ge$_{\rm T}$ (tetrahdral Ge), Sb and Te are depicted with gray, green, yellow and blue spheres.}
    \label{fig:DFT2}
\end{figure}

\begin{figure}[H]
    \centering
           \renewcommand\figurename{Figure~S$\!\!$}
    \includegraphics{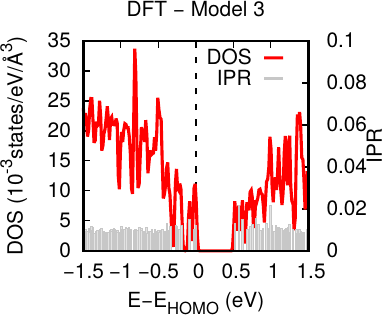}
    \caption{a) A zoom of the electronic DOS of model DFT-3 (red curve) close to the band gap and the value of the IPR (gray spikes) of KS states at the $\Gamma$ point. The zero of energy is the highest occupied state (HOMO). This model dose not display any localized states neither deep nor close to the band edges.}
    \label{fig:DFT3}
\end{figure}

\begin{figure}[H]
    \centering
           \renewcommand\figurename{Figure~S$\!\!$}
    \includegraphics{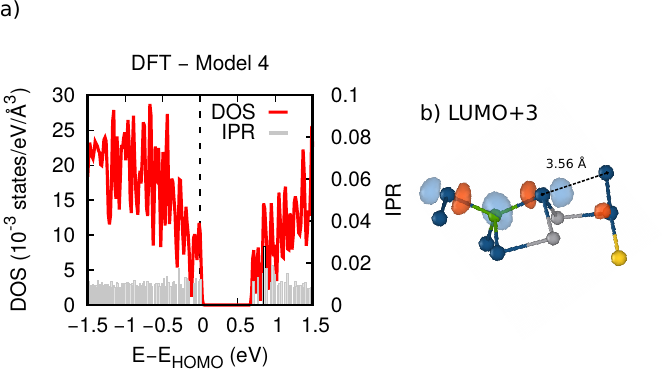}
    \caption{a) A zoom of the electronic DOS of model DFT-4 (red curve) close to the band gap and the value of the IPR (gray spikes) of KS states at the $\Gamma$ point. The zero of energy is the highest occupied state (HOMO).   The isosurfaces of the state with highest IPR (highlighted with black border) is shown in panel b. b) The LUMO+3 state is localized close to a Ge$_{\rm T}$ atom. Ge, Ge$_{\rm T}$ (tetrahdral Ge), Sb and Te are depicted with gray, green, yellow and blue spheres.}
    \label{fig:DFT4}
\end{figure}

\begin{figure}[H]
    \centering
           \renewcommand\figurename{Figure~S$\!\!$}
    \includegraphics{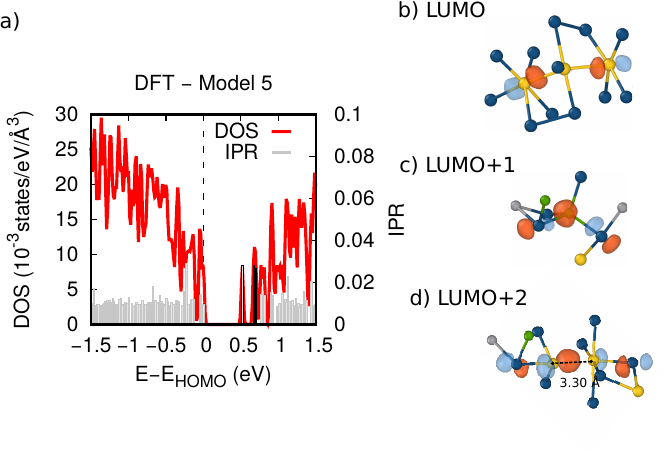}
    \caption{a) A zoom of the electronic DOS of model DFT-5 (red curve) close to the band gap and the value of the IPR (gray spikes) of KS states at the $\Gamma$ point. The zero of energy is the highest occupied state (HOMO). The isosurfaces of states with high IPR (highlighted with black border) are shown in panels b and c. b) The LUMO  is localized on a short chain of Sb atoms.  c) LUMO+1 is localized close to a Ge$_{\rm T}$ atom. d) LUMO+2 is localized close to a 5-coordinated Sb atom forming long contact with a Sb atom. Ge, Ge$_{\rm T}$ (tetrahdral Ge), Sb and Te are depicted with gray, green, yellow and blue spheres.}
    \label{fig:DFT5}
\end{figure}

\begin{figure}[H]
    \centering
           \renewcommand\figurename{Figure~S$\!\!$}
    \includegraphics{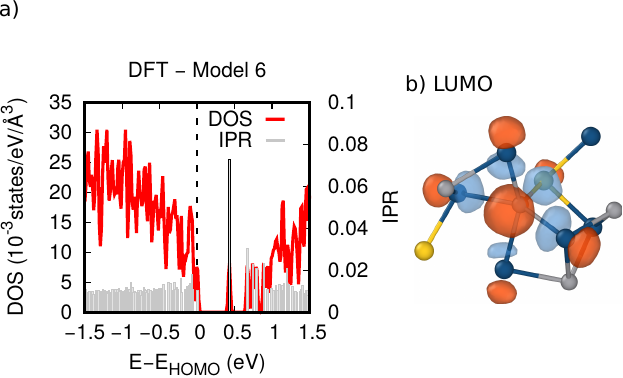}
    \caption{a) A zoom of the electronic DOS of model DFT-6 (red curve) close to the band gap and the value of the IPR (gray spikes) of KS states at the $\Gamma$ point. The zero of energy is the highest occupied state (HOMO).   The isosurfaces of the state with highest IPR (highlighted with black border) is shown in panel b. b) The LUMO state is localized close to a  5-coordinated Ge atom. Ge, Ge$_{\rm T}$ (tetrahdral Ge), Sb and Te are depicted with gray, green, yellow and blue spheres.}
    \label{fig:DFT6}
\end{figure}

\begin{figure}[H]
    \centering
           \renewcommand\figurename{Figure~S$\!\!$}
    \includegraphics{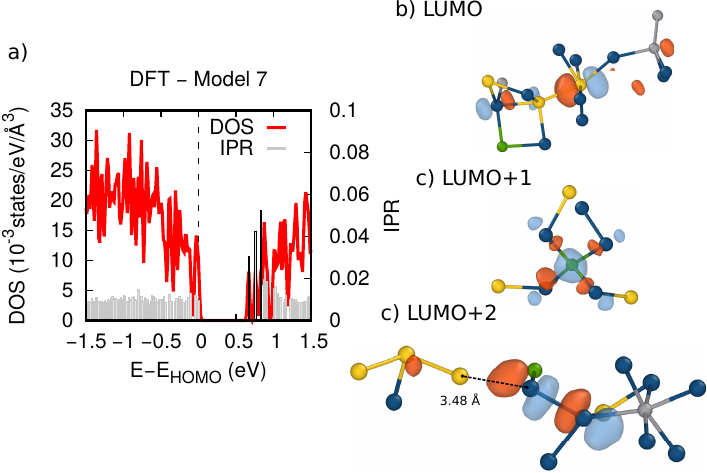}
    \caption{a) A zoom of the electronic DOS of model DFT-7 (red curve) close to the band gap and the value of the IPR (gray spikes) of KS states at the $\Gamma$ point. The zero of energy is the highest occupied state (HOMO). The isosurfaces of states with high IPR (highlighted with black border) are shown in panels b and c. b) The LUMO is localized on a homopolar Sb-Sb bond.  c) LUMO+1 is localized an a Ge$_{\rm T}$ atom. d) LUMO+2 is localized on a Te-Te homopolar bond. Ge, Ge$_{\rm T}$ (tetrahdral Ge), Sb and Te are depicted with gray, green, yellow and blue spheres.}
    \label{fig:DFT7}
\end{figure}

\begin{figure}[H]
    \centering
           \renewcommand\figurename{Figure~S$\!\!$}
    \includegraphics{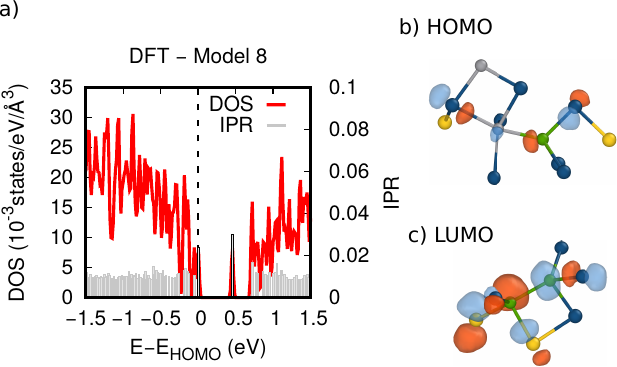}
    \caption{a) A zoom of the electronic DOS of model DFT-8 (red curve) close to the band gap and the value of the IPR (gray spikes) of KS states at the $\Gamma$ point. The zero of energy is the highest occupied state (HOMO). The isosurfaces of states with high IPR (highlighted with black border) are shown in panels b and c. b) The HOMO  is localized on a homopolar Ge-Ge$_{\rm T}$ bond.  c) LUMO is localized close to two bonded Ge$_{\rm T}$ atoms. Ge, Ge$_{\rm T}$ (tetrahdral Ge), Sb and Te are depicted with gray, green, yellow and blue spheres.}
    \label{fig:DFT8}
\end{figure}

\begin{figure}[H]
    \centering
           \renewcommand\figurename{Figure~S$\!\!$}
    \includegraphics{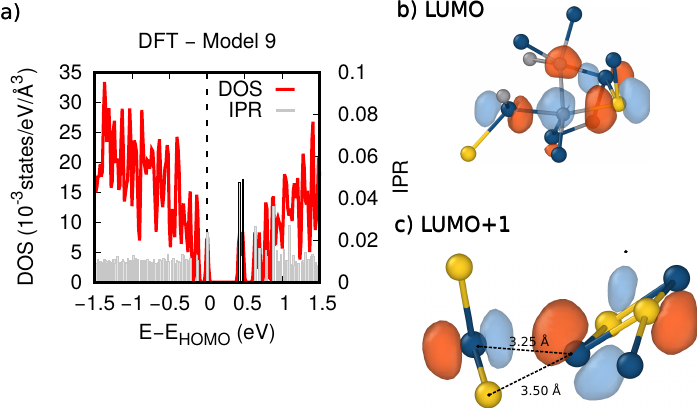}
    \caption{a) A zoom of the electronic DOS of model DFT-9 (red curve) close to the band gap and the value of the IPR (gray spikes) of KS states at the $\Gamma$ point. The zero of energy is the highest occupied state (HOMO). The isosurfaces of states with high IPR (highlighted with black border) are shown in panels b and c. b) The LUMO  is localized on a 5-coordinated Ge atom forming a bond with undercoordinated Sb atom and another 5-coordinated Ge atom.  c) LUMO+1 is localized on two Te atoms with long Te-Te and Te-Sb contacts. Ge, Ge$_{\rm T}$ (tetrahdral Ge), Sb and Te are depicted with gray, green, yellow and blue spheres.}
    \label{fig:DFT9}
\end{figure}

\begin{figure}[H]
    \centering
           \renewcommand\figurename{Figure~S$\!\!$}
    \includegraphics{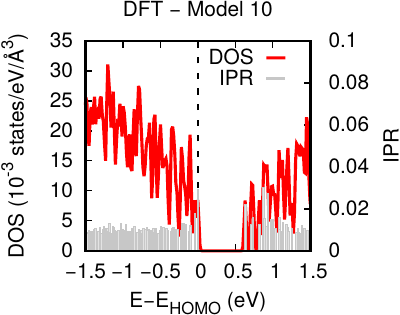}
    \caption{a) A zoom of the electronic DOS of model DFT-10 (red curve) close to the band gap and the value of the IPR (gray spikes) of KS states at the $\Gamma$ point. The zero of energy is the highest occupied state (HOMO). This model dose not display any localized states neither deep nor close to the band edges.}
    \label{fig:DFT10}
\end{figure}  

\section{Electronic DOS of 459-atom models}
In Figs. S41-S44 we show the electronic density of states (DOS) computed with the HSE hydrid functional \cite{hse06} of the four 459-atom amorphous models of GST generated by quenching the melt in DFT-MD  in Ref. \cite{caravatiblyp} and then geometry-optimized at zero temperature with the HSE functional. Isosurfaces of the most localized in-gap states analyzed in the article are also shown (see Table S2).
\begin{figure}[H]
    \centering
           \renewcommand\figurename{Figure~S$\!\!$}
    \includegraphics{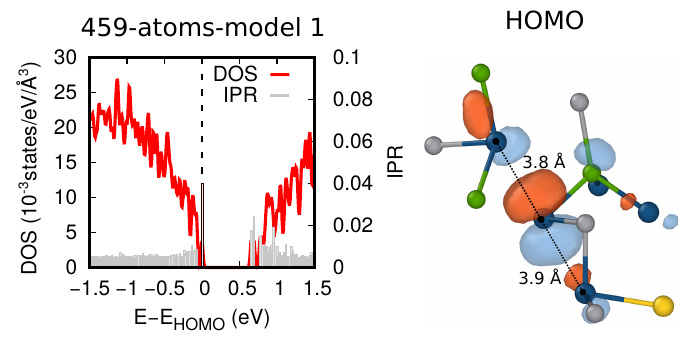}
    \caption{a) A zoom of the electronic DOS of 459-atom model 1 (red curve) close to the
band gap and the value of the IPR (gray spikes) of KS states at the $\Gamma$ point. The zero
energy is the highest occupied state (HOMO). The isosurfaces of states with high IPR
(highlighted with black border) are shown in panels b. b) The HOMO is
localized on a 2-coordinated Te atom forming two long Te-Te contacts. Ge, Ge$_{\rm T}$ (tetrahdral Ge), Sb and Te are depicted with gray, green, yellow and blue
spheres.}
    \label{fig:blyp1}
\end{figure}  

\begin{figure}[H]
    \centering
           \renewcommand\figurename{Figure~S$\!\!$}
    \includegraphics{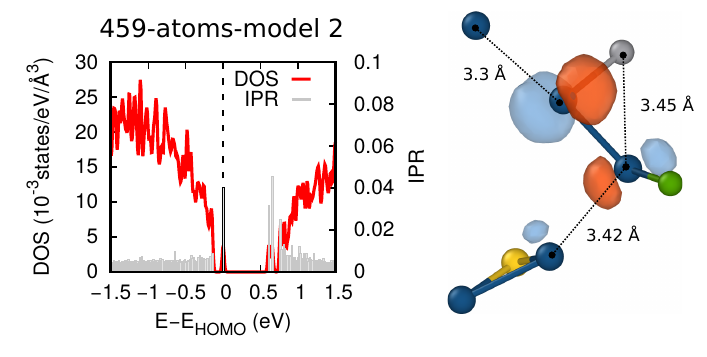}
    \caption{a) A zoom of the electronic DOS of 459-atom model 2 (red curve) close to the
band gap and the value of the IPR (gray spikes) of KS states at the $\Gamma$ point. The zero
energy is the highest occupied state (HOMO). The isosurfaces of states with high IPR
(highlighted with black border) are shown in panels b. b) The HOMO  is
localized on a 2-coordinated Te forming two long Te-Te contacts. Ge, Ge$_{\rm T}$ (tetrahdral Ge), Sb and Te are depicted with gray, green, yellow and blue
spheres.}
    \label{fig:blyp2}
\end{figure}  

\begin{figure}[H]
    \centering
           \renewcommand\figurename{Figure~S$\!\!$}
    \includegraphics{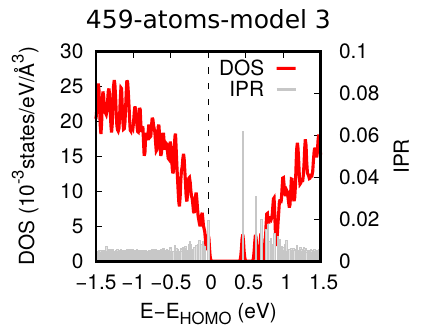}
    \caption{a) A zoom of the electronic DOS of 459-atom model 3 (red curve) close to the
band gap and the value of the IPR (gray spikes) of KS states at the $\Gamma$ point. The zero
energy is the highest occupied state (HOMO).}
    \label{fig:blyp3}
\end{figure}  

\begin{figure}[H]
    \centering
           \renewcommand\figurename{Figure~S$\!\!$}
    \includegraphics{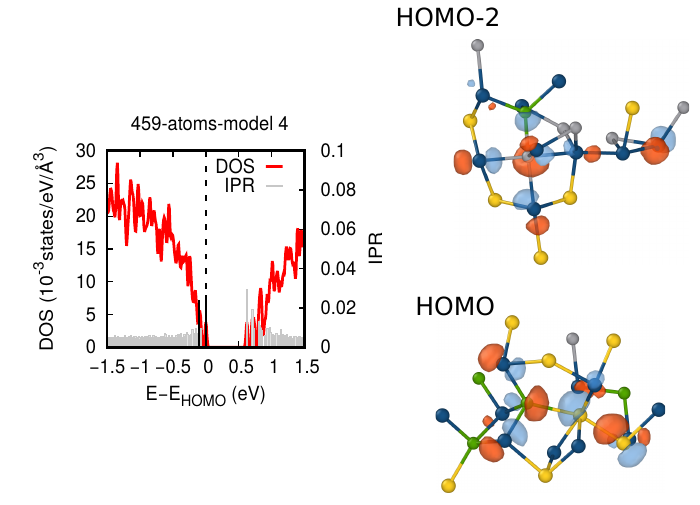}
    \caption{a) A zoom of the electronic DOS of 459-atom model 4 (red curve) close to the
band gap and the value of the IPR (gray spikes) of KS states at the $\Gamma$ point. The zero
energy is the highest occupied state (HOMO). The isosurfaces of states with high IPR
(highlighted with black border) are shown in panels b-c. b) HOMO-2 state is localized on an overcoordinated Ge atom bonded to a short chain of Te atoms. c) The HOMO is
localized on a short chain of wrong bonds involving Sb-Sb and Sb-Ge$_{\rm T}$ bonds. Ge, Ge$_{\rm T}$ (tetrahdral Ge), Sb and Te are depicted with gray, green, yellow and blue
spheres.}
    \label{fig:blyp4}
\end{figure}  

\section{Validation of the electronic properties of NN models}\label{sec:NNvsDFT}
In this section, we report the electronic DOS (see Figs. S45-47) of models generated to benchmark electronic properties of NN models against DFT models.

In Fig. S45 the electronic DOS of four 216-atom NN models (sNN-1, sNN-2, sNN-3 and sNN-4) are compared with the  DOS averaged over the DFT models with the same size.  The DOS of these smaller NN models have no in-gap states in the energy range 0.1-0.3 eV (HOMO at zero energy) as opposed to the larger 297-atom NN models. 

We then verified that the states in the energy range 0.1-0.3 eV observed in the larger NN models are not an artifact of the NN potential, by performing a hybrid quench (NN from 1000 K to 600 K, then DFT down to 300 K) or a hybrid annealing (full quench followed by first annealing with NN-MD, then a second annealing with DFT-MD at 300 K). In Fig. S46 and S47 we report  the electronic DOS of models NN-3 and NN-7 subjected to this protocol that still feature in-gap states in the 0.1-0.3 eV range.

\begin{figure}[H]
    \centering
           \renewcommand\figurename{Figure~S$\!\!$}
       \includegraphics[width=0.8\textwidth, keepaspectratio]{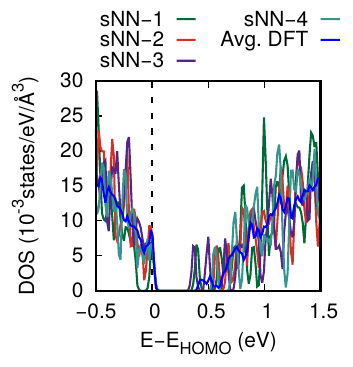}
    \caption{a) A zoom close to the band gap of the  electronic DOS of  sNN models (red curve) compared  with the  DOS  averaged over the DFT models (blue curve).}
    \label{fig:NNvsDFT}
\end{figure}

\begin{figure}[h]
    \centering
           \renewcommand\figurename{Figure~S$\!\!$}
    \includegraphics[width=0.8\textwidth, keepaspectratio]{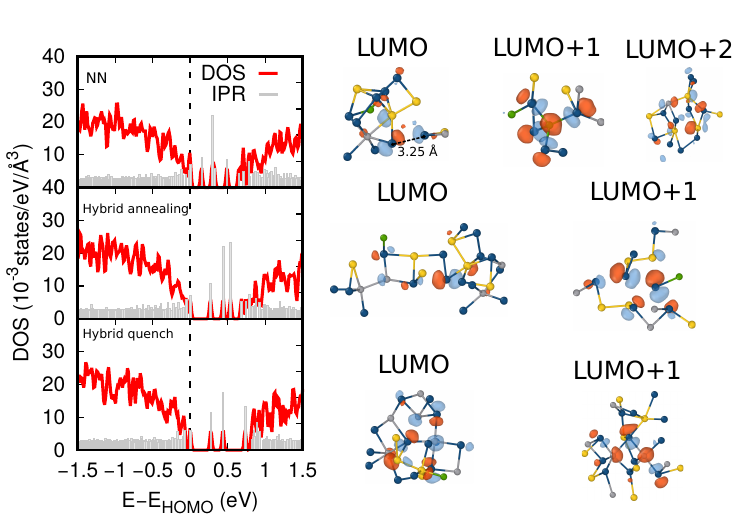}
    \caption{a) A zoom close to the band gap of the  electronic DOS of model NN-3: Full NN quench (upper panel),  hybrid NN+DFT-MD  annealing at 300 K (central panel) and hybrid quench NN+DFT-NN (lower panel, see article). 
    Isosurfaces of in-gap states below 0.35 eV are shown on the left.}
    \label{fig:NN3_3MD}
\end{figure}

\begin{figure}[H]
    \centering
           \renewcommand\figurename{Figure~S$\!\!$}
    \includegraphics{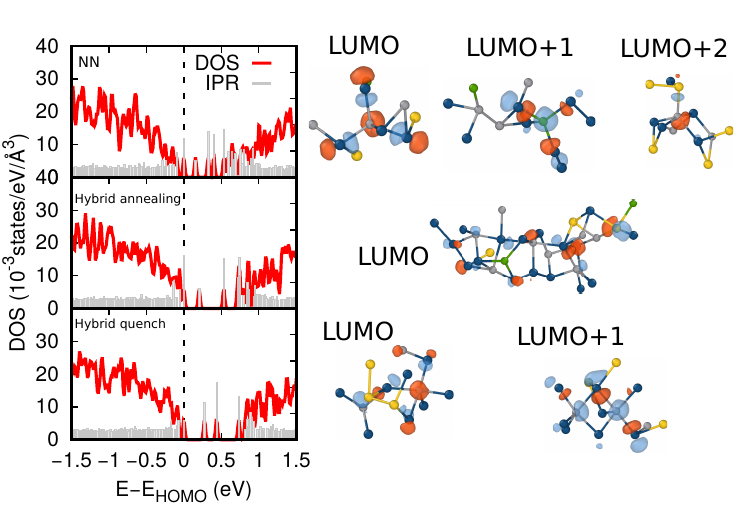}
    \caption{a) A zoom close to the band gap of the  electronic DOS of model NN-7: Full NN quench (upper panel),  hybrid NN+DFT-MD  annealing at 300 K (central panel) and hybrid quench NN+DFT-NN (lower panel, see article).  Isosurfaces of in-gap states below 0.35 eV are shown on the left.}
    \label{fig:NN7_3MD}
\end{figure}

\section{Mobility gap and energy position of in-gap states}

In this section we report  the mobility gap of all 40 models (Table S1) and the position in energy of all in-gap states analyzed in the article (Table S2).

\begin{table}[]
    \centering
               \renewcommand\tablename{Table~S$\!\!$}
        \caption{Mobility gap (eV) of the different models (generated by NN or DFT molecular dynamics) assigned by identifying the
VB edge as the energy below which all states have a IPR lower than 0.018 and are therefore delocalized
and similarly for the CB edges. Due the small size of our simulation cells this procedure is subject to
some arbitrariness and uncertainties.}

    \begin{tabular}{c c c c }
       \hline \hline 
    NN Model    &  Mobility gap (eV)  & DFT Model &  Mobility gap (eV)   \\
     \hline 
NN-1	&	0.74	&	DFT-1	&	0.85	\\
NN-2	&	0.892	&	DFT-2	&	0.74	\\
NN-3	&	0.659	&	DFT-3	&	0.54	\\
NN-4	&	0.81	&	DFT-4	&	0.71	\\
NN-5	&	0.8	&	DFT-5	&	0.77	\\
NN-6	&	0.78	&	DFT-6	&	0.74	\\
NN-7	&	0.78	&	DFT-7	&	0.86	\\
NN-8	&	0.69	&	DFT-8	&	0.76	\\
NN-9	&	0.89	&	DFT-9	&	0.81	\\
NN-10	&	0.96	&	DFT-10	&	0.67	\\
NN-11	&	0.65	&		&		\\
NN-12	&	0.75	&		&		\\
NN-13	&	0.79	&		&		\\
NN-14	&	0.75	&		&		\\
NN-15	&	0.71	&		&		\\
NN-16	&	0.85	&		&		\\
NN-17	&	0.71	&		&		\\
NN-18	&	0.64	&		&		\\
NN-19	&	0.67	&		&		\\
NN-20	&	0.85	&		&		\\
NN-21	&	0.83	&		&		\\
NN-22	&	0.74	&		&		\\
NN-23	&	0.64	&		&		\\
NN-24	&	0.64	&		&		\\
NN-25	&	0.78	&		&		\\
NN-26	&	0.61	&		&		\\
NN-27	&	0.73	&		&		\\
NN-28	&	0.71	&		&		\\
NN-29	&	0.78	&		&		\\
NN-30	&	0.74	&		&		\\

    \hline \hline

    \end{tabular}

\end{table}

\begin{table}[]
    \centering
                  \renewcommand\tablename{Table~S$\!\!$}
        \caption{Position in energy (eV) with respect to the HOMO of the 84 in-gap states in the different models (see text) whose localization properties is analyzed in the article.}

\resizebox{\textwidth}{!}{\begin{tabular}{l c l c l c }
       \hline \hline 
    VB state    &  Energy (eV)  & Deep state    &  Energy (eV)  & CB state    &  Energy (eV)  \\
     \hline 
NN-1 HOMO-2	&	-0.1127	&	NN-1 LUMO	&	0.2743	&	NN-1 LUMO+1	&	0.5383	\\
NN-6 LUMO	&	0.136	&	NN-2 LUMO+4	&	0.3174	&	NN-1 LUMO+2	&	0.6098	\\
NN-7 HOMO	&	0	    &	NN-2 LUMO+5	&	0.6533	&	NN-6 LUMO+1	&	0.6166	\\
NN-10 HOMO	&	0	    &	NN-3 LUMO+1	&	0.2925	&	NN-7 LUMO+3	&	0.5333	\\
NN-17 LUMO	&	0.179	&	NN-4 LUMO	&	0.2401	&	NN-8 LUMO+3	&	0.4997	\\
NN-30 HOMO-1	&	-0.0156	&	NN 4-LUMO+1	&	0.5545	&	NN-8 LUMO+5	&	0.6496	\\
DFT-1 HOMO	&	0	&	NN-5 LUMO	&	0.2957	&	NN-11 LUMO	&	0.3457	\\
DFT-8 HOMO	&	0	&	NN-5 LUMO+2	&	0.4805	&	NN-14 LUMO+2	&	0.6149	\\
459-1 HOMO	&	0	&	NN-7 LUMO+1	&	0.3248	&	NN-16 LUMO+3	&  0.7817	\\
459-2 HOMO	&	0	&	NN-7 LUMO+2	&	0.4009	&	NN-16 LUMO+4	&	0.8126	\\
459-3 HOMO	&	0	&	NN-8 LUMO+2	&	0.4761	&	NN-17 LUMO+2	&	0.517	\\
459-4 HOMO	&	0	&	NN-9 LUMO	&	0.455	&	NN-18 LUMO	&	0.5197	\\
459-4 HOMO-2	&-0.103&	NN-9 LUMO+1	&	0.6188	&	NN-18 LUMO+1	&	0.5861	\\
	&		&	NN-10 LUMO	 &	0.5331	&	NN-19 LUMO	&	0.572	\\
	&		&	NN-12 LUMO	&	0.3598	&	NN-20 LUMO+3	&	0.7166	\\
	&		&	NN-12 LUMO+1	&	0.3679	&	NN-21 LUMO+3	&	0.6809	\\
	&		&	NN-13 LUMO	&	0.4392	&	NN-22 LUMO	&	0.6154	\\
	&		&	NN-14 LUMO	&	0.356	&	NN-22 LUMO+1	&	0.6409	\\
	&		&	NN-14 LUMO+1	&	0.4916	&	NN-23 LUMO+2	&	0.4812	\\
	&		&	NN-15 LUMO	&	0.354	&	NN-26 LUMO+2	&	0.457	\\
	&		&	NN-15 LUMO+1	&	0.4787	&	NN-26 LUMO+3	&	0.5684	\\
	&		&	NN-16 LUMO	&	0.5903	&	NN-27 LUMO+2	&	0.5576	\\
	&		&	NN-20 LUMO	&	0.387	&	NN-27 LUMO+3	&	0.6742	\\
	&		&	NN-20 LUMO+1	&	0.6227	&	NN-30 LUMO	&	0.6164	\\
	&		&	NN-23 LUMO	&	0.3128	&	DFT-1 LUMO	&	0.6917	\\
	&		&	NN-23 LUMO+1	&	0.3905	&	DFT-2 LUMO+1	&	0.6527	\\
	&		&	NN-24 LUMO	&	0.3279	&	DFT-4 LUMO+3	&	0.857	\\
	&		&	NN-24 LUMO+1	&	0.428	&	DFT-5 LUMO+1	&	0.6728	\\
	&		&	NN-25 LUMO	&	0.487	&	DFT-5 LUMO+2	&	0.6975	\\
	&		&	NN-26 LUMO+1	&	0.3461	&	DFT-7 LUMO	&	0.6683	\\
	&		&	NN-27 LUMO	&	0.3267	&	DFT-7 LUMO+1	&	0.7602	\\
	&		&	NN-27 LUMO+1	&	0.4952	&	DFT-7 LUMO+2	&	0.8316	\\
	&		&	NN-28 LUMO	&	0.355	&		&		\\
	&		&	NN-29 LUMO+2	&	0.474	&		&		\\
	&		&	DFT-2 LUMO	&	0.5081	&		&		\\
	&		&	DFT-5 LUMO	&	0.5194	&		&		\\
	&		&	DFT-6 LUMO	&	0.4288	&		&		\\
	&		&	DFT-8 LUMO	&	0.4562	&		&		\\
	&		&	DFT-9 LUMO	&	0.427	&		&		\\
	&		&	DFT-9 LUMO+1	&	0.4732	&		&		\\

    \hline \hline

    \end{tabular}}

\end{table}
\newpage
\bibliographystyle{elsarticle-num} 
\bibliography{ref.bib}